\PassOptionsToPackage{table}{xcolor}
\documentclass[manuscript,screen,research,nonacm]{acmart}

\usepackage[table]{xcolor}
\usepackage{xcolor}

\usepackage{subcaption}
\usepackage{color}
\usepackage{tikz}
\usepackage{forest}
\usepackage{url}
\usepackage{hyperref}
\usepackage{booktabs}
\usepackage{threeparttable}
\usepackage{pifont}
\usepackage[most]{tcolorbox}
\usepackage{multirow}
\tcbuselibrary{breakable}
\usetikzlibrary{positioning,shapes,shadows,arrows}
\usepackage{wasysym}
\usepackage{adjustbox}
\definecolor{ngreen}{HTML}{D5E8D4}
\definecolor{nblue}{HTML}{DAE8FC}
\definecolor{npurple}{HTML}{E1D5E7}

\newcommand{\lj}{\left \langle}
\newcommand{\rj}{\right \rangle}

\newcommand{\tabincell}[2]{\begin{tabular}{@{}#1@{}}#2\end{tabular}}
\usepackage{array} 
\definecolor{myblue}{RGB}{204,217,245}

\AtBeginDocument{%
  }

\author{Wenxuan Zeng}
\authornote{Authors contributed equally.}
\author{Tianshi Xu}
\authornotemark[1]
\author{Yi Chen}
\authornotemark[1]
\author{Yifan Zhou}
\affiliation{
    \institution{Peking University}
    \country{China}
}
\author{Mingzhe Zhang, Jin Tan, Cheng Hong}
\affiliation{
    \institution{Ant Group}
    \country{China}
}
\author{Meng Li}
\authornote{Corresponding author (meng.li@pku.edu.cn).}
\affiliation{
    \institution{Peking University}
    \country{China}
}

\begin{document}

\title{Towards Efficient Privacy-Preserving Machine Learning: A Systematic Review from Protocol, Model, and System Perspectives}




\renewcommand{\shortauthors}{Wenxuan Zeng et al.}

\begin{abstract}

    Privacy-preserving machine learning (PPML) based on cryptographic protocols has emerged as a promising paradigm to protect user data privacy in cloud-based machine learning services. While it achieves formal privacy protection, PPML often incurs significant efficiency and scalability costs due to orders of magnitude overhead compared to the plaintext counterpart. 
    Therefore, there has been a considerable focus on mitigating the efficiency gap for PPML. In this survey, we provide a comprehensive and systematic review of recent PPML studies with a focus on cross-level optimizations. Specifically, we categorize existing papers into protocol level, model level, and system level, and review progress at each level. We also provide qualitative and quantitative comparisons of existing works with technical insights, based on which we discuss future research directions and highlight the necessity of integrating optimizations across protocol, model, and system levels. We hope this survey can provide an overarching understanding of existing approaches and potentially inspire future breakthroughs in the PPML field. As the field is evolving fast, we also provide a public GitHub repository to continuously track the developments, which is available at \url{https://github.com/PKU-SEC-Lab/Awesome-PPML-Papers}.

\end{abstract}

\ccsdesc[500]{General and reference~Surveys and overviews}
\ccsdesc[500]{Security and privacy~Privacy-preserving protocols}
\ccsdesc[500]{Computing methodologies~Artificial intelligence}

\maketitle


\section{Introduction}


With the advent of machine learning (ML), artificial intelligence (AI) has ushered in an unprecedented era, profoundly benefiting diverse aspects of society such as smart homes \cite{sepasgozar2020systematic,almusaed2023enhancing}, intelligent manufacturing \cite{nti2022applications,li2017applications}, and smart healthcare \cite{gao2024artificial,sujith2022systematic}.
In recent years, Transformer-based models \cite{vaswani2017attention}, especially large language models (LLMs), have emerged as a game-changing revolution in the AI field, such as ChatGPT \cite{chatgpt}, DeepSeek \cite{liu2024deepseek,guo2025deepseek}, and Claude \cite{claude4}.
These models demonstrate advanced capabilities in multimodal understanding and complex reasoning \cite{huang2022towards,wei2022chain,chen2025towards}.

While the models demonstrate exceptional performance, ML as a service (MLaaS) on the cloud has raised serious privacy issues \cite{rathee2020cryptflow2,rathee2021sirnn,mishra2020delphi,huang2022cheetah}.
The clients are required to upload their input prompts to the cloud, which may contain sensitive personal information. Meanwhile, the service provider (i.e., server) like OpenAI is unwilling to offload the trained model to the user in order to protect the proprietary model weights.
Hence, despite the convenience and computational power offered by MLaaS, privacy issues remain a critical obstacle to the broader deployment of AI in real-world settings.

\begin{figure}
    \centering
    \includegraphics[width=\linewidth]{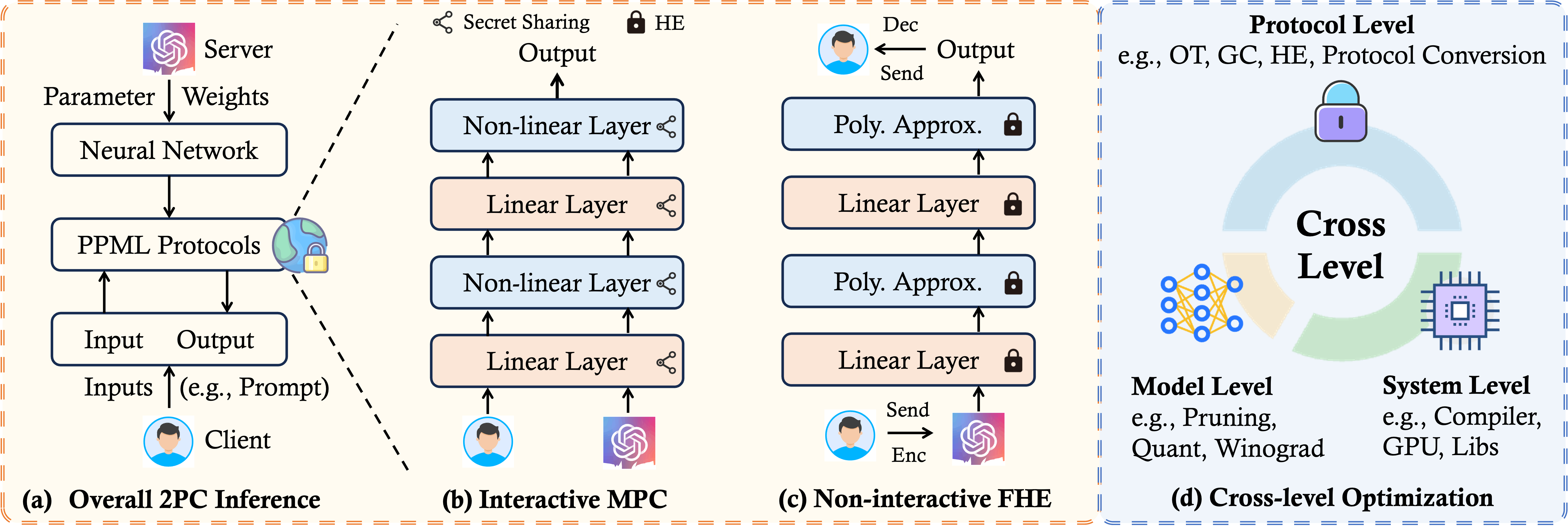}
    \caption{(a) Overall 2PC inference framework, involving a client and a server. (b) Interactive MPC inference with secret sharing. (c) Non-interactive FHE inference with polynomial approximation for non-linear layers. (d) Paradigm of cross-level optimization for PPML, including protocol, model, and system.}
    \label{fig:2pc_framework}
\end{figure}

To address the issue, privacy-preserving machine learning (PPML) has become a promising and prevalent paradigm for cryptographically strong data privacy protection, fulfilling both parties’ requirements\footnote{This paper primarily focuses on two-party computation (2PC) while computation involving three or more parties will only be briefly discussed.}: 
the server learns nothing about the user's input, and the user learns nothing about the server's weights, apart from the final inference results as shown in Figure \ref{fig:2pc_framework}(a).
PPML includes many research areas such as multi-party computation (MPC), fully homomorphic encryption (FHE), differential privacy (DP), trusted execution environment (TEE), federated learning (FL), etc. In this survey, we narrow down the discussed topic to MPC and FHE, and focus on the optimizations across the protocol level, model level, and system level.
PPML can be categorized into interactive MPC inference and non-interactive FHE inference as shown in Figure \ref{fig:2pc_framework}.
For MPC as shown in Figure \ref{fig:2pc_framework}(b), the client and server jointly perform the linear and non-linear layer computation using the secret sharing (SS) scheme.
Commonly, non-linear layers are evaluated using garble circuit (GC) \cite{mishra2020delphi,riazi2019xonn,jha2021deepreduce} or oblivious transfer (OT) \cite{rathee2020cryptflow2,rathee2021sirnn,zeng2023copriv} while linear layers are evaluated using OT \cite{rathee2020cryptflow2,rathee2021sirnn,zeng2023copriv} or homomorphic encryption (HE) \cite{huang2022cheetah,juvekar2018gazelle,xu2024privcirnet}.
For FHE as shown in Figure \ref{fig:2pc_framework}(c), the client first encrypts the input and then sends it to the server.
The server computes the entire model without decryption and then sends the final encrypted output to the client for decryption.

However, the formal privacy protection of PPML comes at a significant
cost in terms of efficiency and scalability, as it often incurs orders of magnitude overhead compared to the plaintext counterpart.
In the past few years, numerous optimizations have been proposed to achieve better efficiency at the protocol level, model level, and system level.
However, there is still a lack of surveys that systematically analyze existing studies and discuss future research directions in the PPML field.
In this survey, we highlight three opportunities at the following three levels: 
\begin{itemize}
    \item \textbf{Opportunity 1 (Protocol Level):} There exists a significant gap between the PPML protocol and plaintext computation. For example, OT brings huge communication overhead, and 
    HE brings huge computational overhead as it operates on ciphertexts representing high-dimensional data but with limited vectorization capabilities (e.g., one-dimensional encoding).
    Different protocols, such as HE encoding methods, directly affect the overhead. Moreover, existing protocol designs often overlook the inherent characteristics of models, e.g., the robustness to quantization noise and model sparsity. 
    These observations point to the \textit{\textbf{opportunity for the protocol-level optimization itself and model-aware optimization.}}
    \item \textbf{Opportunity 2 (Model Level):} Existing model architectures are not friendly to PPML, and model-level optimizations designed for plaintext overlook the characteristics of the underlying protocols. For example, 
    ReLU introduces significant communication cost compared to the plaintext counterpart. However, 
    directly pruning ReLUs 
    ignores the overhead of the linear layer and truncation protocols. Moreover, quantization introduces extra online costs like bit width extension and truncation. 
    Therefore, we highlight the \textit{\textbf{opportunity for PPML-friendly model architecture design and protocol-aware optimization.}}
    \item \textbf{Opportunity 3 (System Level):} 
    Current system-level optimizations are often developed independently of the underlying protocol characteristics, leaving room for tighter integration. For example, many compilers assume fixed encoding domains and lack support for diverse protocol features. Similarly, GPU-based methods typically target general-purpose acceleration but overlook workload-specific factors such as batch size sensitivity and evaluation key usage. These observations point to a valuable \textit{\textbf{opportunity for protocol-aware system design, which can significantly improve the efficiency and scalability of PPML deployments.}}
\end{itemize}

Considering these three opportunities,
in this survey, we comprehensively review the PPML studies across the protocol level, model level, and system level as shown in Figure \ref{fig:2pc_framework}(d).
In contrast to our survey, the prior PPML survey \cite{li2024survey} 
does not differentiate between protocol-level and model-level optimizations and lacks system-level discussions.
\cite{chen2025towards} and \cite{ng2023sok} lack discussions about model-level optimizations.
None of these surveys bridges the gap across the three levels or explores the relationship, challenges, and opportunities among them. 
Therefore, we position this survey as the first to systematically review the PPML field across the protocol level, model level, and system level, with a meticulous PPML taxonomy.
To summarize, we make the following contributions in this survey:
\begin{itemize}
    \item 
    We conduct a comprehensive survey across three levels:
    \textbf{1) protocol level:} we primarily review OT-based and HE-based protocols, identifying the most suitable protocols and HE encoding methods for different scenarios. \textbf{2) model level:} we review the studies of linear layer, non-linear layer, and quantization, highlighting the importance of simultaneously optimizing all layers and also emphasizing the necessity of protocol-aware model-level optimization; \textbf{3) system level:} we review recent advances in compiler and GPU optimizations, emphasizing the need for PPML-friendly system designs and clarifying how system components can align with protocol characteristics to improve efficiency and scalability.
    \item 
    We provide qualitative and quantitative comparisons of existing works with technical insights at the protocol level, model level, and system level.
    Moreover, we provide technical guidelines and discuss the promising avenues for future research, including the \textit{\textbf{necessity of integrating optimizations across the protocol level, model level, and system level.}}
    \item 
    To continuously track the developments in the PPML field, we also provide a public GitHub repository, which is available at \url{https://github.com/PKU-SEC-Lab/Awesome-PPML-Papers}. 
\end{itemize}

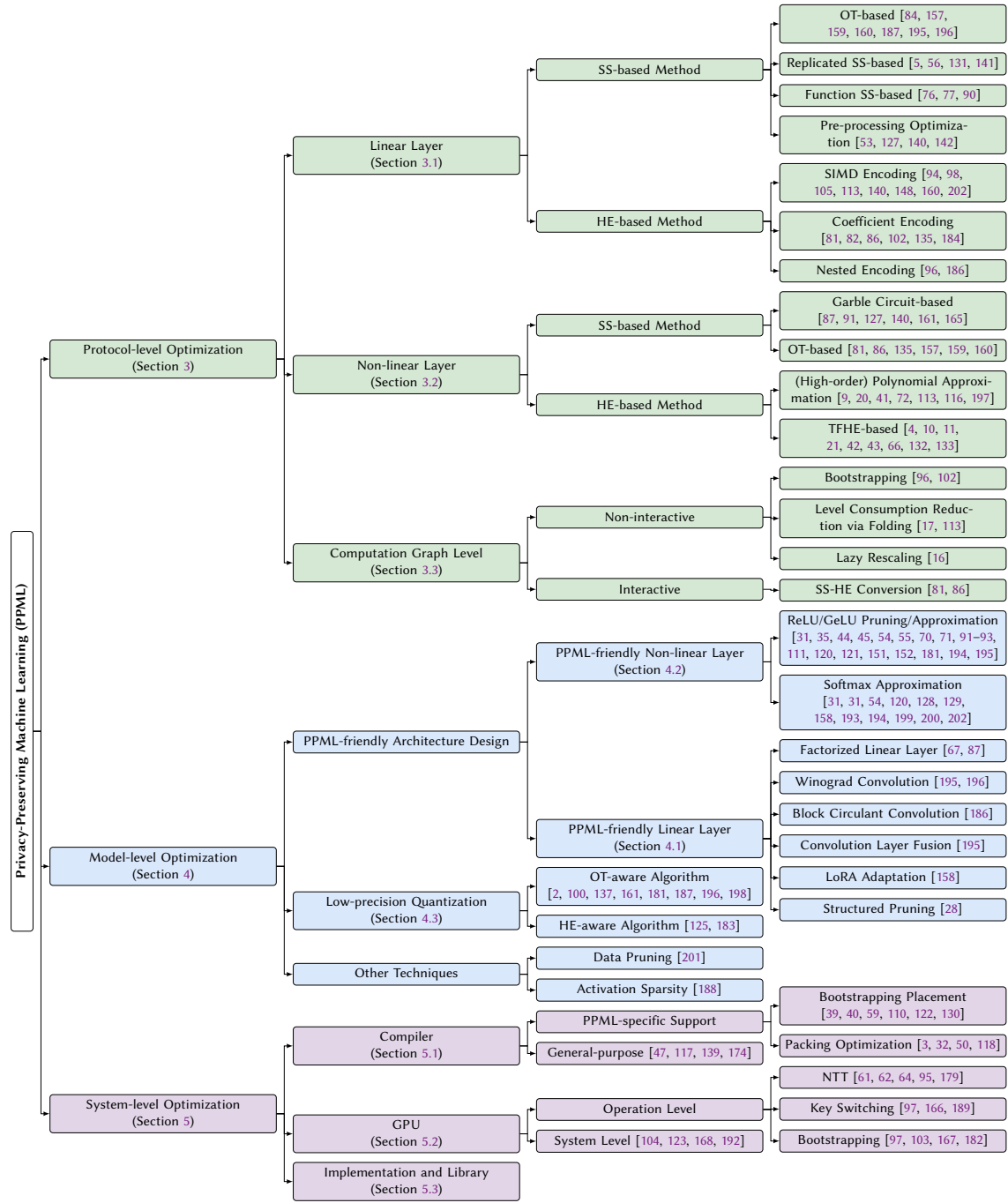
\begin{figure}[!tb]
\centering
\tikzset{
    basic/.style  = {draw, text width=2cm, align=center, font=\sffamily, rectangle},
    root/.style   = {basic, rounded corners=2pt, thin, align=center, fill=white!20, text width=9cm, rotate=90, font=\sffamily, minimum height=0.5cm},
    dnode/.style = {basic, thin, rounded corners=2pt, align=center, fill=ngreen, text width=5cm, font=\sffamily, minimum height=0.5cm},
    dnode_1/.style = {basic, thin, rounded corners=2pt, align=center, fill=ngreen,text width=5cm, font=\sffamily, minimum height=0.5cm},
    mnode/.style = {basic, thin, rounded corners=2pt, align=center, fill=nblue, text width=5cm, font=\sffamily, minimum height=0.5cm},
    mnode_1/.style = {basic, thin, rounded corners=2pt, align=center, fill=nblue, text width=5cm, font=\sffamily, minimum height=0.5cm}, 
    snode/.style = {basic, thin, rounded corners=2pt, align=center, fill=npurple, text width=5cm, font=\sffamily, minimum height=0.5cm},
    snode_1/.style = {basic, thin, rounded corners=2pt, align=center, fill=npurple,text width=5cm, font=\sffamily, minimum height=0.5cm},
}
\resizebox{\linewidth}{!}{
\begin{forest} 
for tree={
    if level=0{
        grow=east,
        growth parent anchor=east,
        parent anchor=south,
        child anchor=west,
        edge path={\noexpand\path[\forestoption{edge},->, >={latex}] 
             (!u.parent anchor) -- +(5pt,0pt) |- (.child anchor)
             \forestoption{edge label};},
    }
    {
        grow=east,
        growth parent anchor=east,
        parent anchor=east,
        child anchor=west,
        edge path={\noexpand\path[\forestoption{edge},->, >={latex}] 
             (!u.parent anchor) -- +(5pt,0pt) |- (.child anchor)
             \forestoption{edge label};},
    }
}
[\textbf{Privacy-Preserving Machine Learning (PPML)}, root
    [System-level Optimization \\ (Section \ref{sec:system}), snode_1
        [Implementation and Library \\ (Section \ref{sec:lib}), snode_1]
        [GPU \\ (Section \ref{sec:gpu}) , snode_1
            [System Level \cite{shivdikar2023gme,yudha2024boostcom,li2025cat,kim2025anaheim}, snode_1]
            [Operation Level, snode_1
                [Bootstrapping \cite{jung2021over,kim2024cheddar,xiao2025gpu,shen2025velofhe}, snode]
                [Key Switching \cite{jung2021over,shen2022carm,yang2024phantom}, snode]
                [NTT \cite{wang2023he,fan2023towards,fan2023tensorfhe,fan2025warpdrive,jiao2025neo}, snode]
            ]
        ]
        [Compiler \\ (Section \ref{sec:compiler}), snode_1
            [General-purpose \cite{cowan2021porcupine,lee2023elasm,malik2023coyote,viand2023heco}, snode_1]
            [PPML-specific Support, snode_1
                [Packing Optimization \cite{dathathri2019chet,chen2020ahec,lee2022hecate,aharoni2020helayers}, snode]
                [Bootstrapping Placement \cite{krastev2024tensor,cheon2024dacapo,cheon2025halo,liu2025resbm,li2025ant,ebel2023orion}, snode]
            ]
        ]
    ]
    [Model-level Optimization \\ (Section \ref{sec:model}), mnode_1
        [Other Techniques, mnode_1 
            [Activation Sparsity \cite{yan2025comet}, mnode_1]
            [Data Pruning \cite{zhangheprune}, mnode_1]
        ]
        [Low-precision Quantization \\ (Section \ref{sec:quant}), mnode_1
            [HE-aware Algorithm \cite{lin2024fastquery,xu2024hequant}, mnode_1]
            [OT-aware Algorithm \cite{xu2024privquant,zeng2024eqo,wu2024ditto,luo2023aq2pnn,agrawal2019quotient,riazi2019xonn,zhang2024scalable,keller2022secure}, mnode_1]  
        ]
        [PPML-friendly Architecture Design, mnode_1
            [PPML-friendly Linear Layer \\ (Section \ref{sec:model_linear}), mnode_1
                [Structured Pruning \cite{hunter}, mnode]
                [LoRA Adaptation \cite{rathee2024mpc}, mnode]
                [Convolution Layer Fusion \cite{zeng2023copriv}, mnode]
                [Block Circulant Convolution \cite{xu2024privcirnet}, mnode]
                [Winograd Convolution \cite{zeng2023copriv,zeng2024eqo}, mnode]
                [Factorized Linear Layer \cite{hussain2021coinn,ganesan2022efficient}, mnode]
            ]
            [PPML-friendly Non-linear Layer \\ (Section \ref{sec:model_nonlinear}), mnode_1
                [Softmax Approximation \cite{zeng2023mpcvit,li2022mpcformer,zhang2023sal,dhyani2023privit,chen2023rna,zimerman2024power,liu2023llms,liu2025mlformer,chen2023rna,rathee2024mpc,zhang2025cipherprune,zeng2025mpcache}, mnode]
                [ReLU/GeLU Pruning/Approximation \cite{zeng2023mpcvit,li2022mpcformer,cho2022selective,jha2021deepreduce,kundu2023learning,liseesaw,park2022aespa,ghodsi2021circa,peng2023autorep,jha2024aero,diaa2024fast,zeng2023copriv,10224651,ghodsi2020cryptonas,chen2023rna,wu2024ditto,jha2024deepreshaperedesigningneuralnetworks,cho2021sphynx,dhyani2023privit}, mnode]
            ]
        ]
    ]
    [Protocol-level Optimization \\ (Section \ref{sec:protocol}), dnode_1
        [Computation Graph Level \\ (Section \ref{sec:graph}), dnode_1
            [Interactive, dnode_1
                [SS-HE Conversion \cite{hao2022iron,huang2022cheetah}, dnode]
            ]
            [Non-interactive, dnode_1
                [Lazy Rescaling \cite{boemer2019ngraph2}, dnode]
                [Level Consumption Reduction via Folding \cite{boemer2019ngraph,lee2022low}, dnode]
                [Bootstrapping \cite{kim2023optimized,ju2024neujeans}, dnode]
            ]
        ]
        [Non-linear Layer \\ (Section \ref{sec:protocol_nonlinear}), dnode_1
            [HE-based Method, dnode_1
                [TFHE-based \cite{chillotti2020tfhe,frery2023privacy,boura2019simulating,chillotti2021programmable,zama2024TFHE,liu2023amortized,liu2025relaxed,bae2024bootstrapping,bae2025bootstrapping}, dnode]
                [(High-order) Polynomial Approximation \cite{zhang2024secure,lee2022privacy,gilad2016cryptonets,ao2024autofhe,boura2018chimera,chialva2018conditionals,lee2022low}, dnode]
            ]
            [SS-based Method, dnode_1
                [OT-based \cite{rathee2020cryptflow2,rathee2021sirnn,rathee2022secfloat,huang2022cheetah,hao2022iron,lu2023bumblebee}, dnode]
                [Garble Circuit-based \cite{jha2021deepreduce,mishra2020delphi,hussain2021coinn,shen2022abnn2,riazi2019xonn,liu2017oblivious}, dnode]
            ]
        ]
        [Linear Layer \\ (Section \ref{sec:protocol_linear}), dnode_1
            [HE-based Method, dnode_1
                [Nested Encoding \cite{ju2024neujeans,xu2024privcirnet}, dnode]
                [Coefficient Encoding \cite{huang2022cheetah,hao2022iron,xu2023falcon,lu2023bumblebee,he2024rhombus,kim2023optimized}, dnode]
                [SIMD Encoding \cite{juvekar2018gazelle,rathee2020cryptflow2,kim2022secure,lee2022low,mishra2020delphi,pang2024bolt,jiang2018secure,zimerman2024power}, dnode]
            ]
            [SS-based Method, dnode_1
                [Pre-processing Optimization \cite{mishra2020delphi,liu2017oblivious,mohassel2017secureml,demmler2015aby}, dnode]
                [Function SS-based \cite{jawalkar2024orca,gupta2023sigma,gupta2022llama}, dnode]
                [Replicated SS-based \cite{mohassel2018aby3,dong2023puma,akimoto2023privformer,liu2024pptif}, dnode]
                [OT-based \cite{rathee2020cryptflow2,rathee2021sirnn,rathee2022secfloat,zeng2023copriv,xu2024privquant,hou2023ciphergpt,zeng2024eqo}, dnode]
            ]
        ]
    ]
]
\end{forest}
}
\caption{Taxonomy of existing PPML studies, including protocol-level, model-level, and system-level optimizations.}
\label{fig:overall}
\end{figure}

\textbf{Taxonomy and Paper Organization.}
The taxonomy of PPML is shown in Figure \ref{fig:overall}.
We categorize PPML into protocol-level optimization (Section \ref{sec:protocol}), model-level optimization (Section \ref{sec:model}), and system-level optimization (Section \ref{sec:system}).
We divide the protocol-level optimization into linear layers (Section \ref{sec:protocol_linear}), non-linear layers (Section \ref{sec:protocol_nonlinear}), and computation graph (Section \ref{sec:graph}).
At the model level, we focus on PPML-friendly model architecture design for linear layer (Section \ref{sec:model_linear}), non-linear layer (Section \ref{sec:model_nonlinear}), and quantization designed for OT and HE protocols (Section \ref{sec:quant}).
We discuss system-level optimization from the compiler (Section \ref{sec:compiler}) and GPU tailored for PPML (Section \ref{sec:gpu}).
Finally, we provide an in-depth discussion of the key challenges in cross-level optimization and also provide our view of PPML in the era of LLMs.

\section{Background and Preliminaries}

\subsection{Neural Networks in Machine Learning}


\subsubsection{Convolutional Neural Networks}

Convolutional Neural Networks (CNNs) mainly involve convolution layers and non-linear activation layers, including rectified linear unit (ReLU) and pooling layers.

\textbf{Convolution layer.}
Convolution layers are the core building blocks of CNN and involve a lot of multiplications.
The parameters consist of a set of learnable filters (or kernels).
A single channel 2-dimensional convolution between the filter weights $W$ and the input $X$ can be formulated as
\begin{equation}
    \mathrm{Conv}_{i,j} = \sum_{m=0}^{M-1}\sum_{n=0}^{N-1} W_{m,n} \cdot X_{i+m,j+n},
\end{equation}
where $M, N$ are the height and width of the kernel, respectively.

\textbf{ReLU}.
ReLU is the most common non-linear layer defined as $\mathrm{ReLU}(x) = \max(0, x)$, which requires element-wise comparison operations and is not cheap in PPML.

\textbf{Pooling.}
Pooling layers are used to reduce the shape of feature maps by combining the feature maps into clusters.
Max pooling and average pooling are widely used, where the former computes the maximum value per cluster in the feature map and the latter takes the average value.

\subsubsection{Transformer-Based Models}


The basic Transformer architecture consists of an encoder and a decoder \cite{vaswani2017attention}.
Vision Transformers (ViTs) follow encoder-only architecture \cite{dosovitskiy2020image} while
auto-regressive generative LLMs follow decoder-only architecture \cite{radford2019language}.
Here, we mainly focus on introducing the decoder-only architecture, and other Transformer-based models share a similar design.
An LLM consists of two modules, i.e., attention and feed-forward network (FFN).


\textbf{Attention.}
Attention utilizes the dot product to effectively capture the relationship and dependency among the tokens.
Attention is defined as
\begin{equation}
    \mathrm{Attention}(Q, K, V) = \mathrm{Softmax}({QK^T}/{\sqrt{d_k}})\cdot V,
\end{equation}
where $d_k$ is the embedding dimension of the key.
Multi-head attention (MHA) is further proposed to guide the model to jointly attend to information from different representation subspaces at different positions, i.e., attention heads.

The generative LLM inference procedure is generally in an autoregressive style such as GPT-2 \citep{radford2019language} and LLaMA \citep{touvron2023llama}, and mainly
consists of two stages: 1) the prefill (prompt) stage and 2) the decoding (generation) stage.

\textit{1) Prefill stage.}
The prefill stage serves as the first step of generation.
LLM takes a prompt sequence as input and generates key-value (KV) caches for each layer. The attention can be computed as
$
    \mathbf O_{\texttt{prompt}} = \mathrm{Softmax}(\mathbf Q_{\texttt{prompt}} \cdot \mathbf K_{\texttt{prompt}}^\top / \sqrt{d})\cdot \mathbf V_{\texttt{prompt}},
$
where $\mathbf Q_{\texttt{prompt}}\in\mathbb R^{H\times T\times d}$ denotes the query and $\mathbf K_{\texttt{prompt}}\in\mathbb R^{H\times T\times d}, \mathbf V_{\texttt{prompt}}\in\mathbb R^{H\times T\times d}$ denote the key and value cache, respectively.
After the prefill stage, the KV cache is generated,
which is the foundation for the downstream decoding stage.
KV cache retains previously computed key-value pairs, eliminating the need for costly re-computation of previous key and value vectors.

\textit{2) Decoding stage.}
The decoding stage uses and updates the stored KV cache to generate new tokens step by step.
Therefore, the attention can be computed as
$
    \mathbf o_{\texttt{dec}} = \mathrm{Softmax}(\mathbf q_{\texttt{dec}} \cdot \mathbf K_{\texttt{cache}}^\top / \sqrt{d})\cdot \mathbf V_{\texttt{cache}},
$
where $\mathbf q_{\texttt{dec}}\in\mathbb R^{1\times d}$ denotes the current query.
The attention output $\mathbf o_{\texttt{dec}}\in\mathbb R^{1\times d}$ is then sent to the FFN layer for subsequent computation.

\textbf{FFN.}
For widely used designs like GPT-2 \cite{radford2019language}, FFN consists of two linear layers and a Gaussian Error Linear Unit (GeLU) non-linear activation function, which can be formulated as
\begin{equation}
    \mathrm{FFN}(x) = W_2(\mathrm{GeLU}(W_1x + b_1)) + b_2,
\end{equation}
where $W_1, W_2$ are linear transformations, and $b_1, b_2$ are two biases.
For LLaMA models, FFN is more complex with the SwiGLU activation function and is defined as
\begin{equation}
    \mathrm{FFN}(x) = W_2(\mathrm{SiLU}(W_gx)\odot (W_1x)),
\end{equation}
where $W_1, W_2, W_g$ are linear transformations, $\odot$ is element-wise multiplication.

\subsection{PPML Primitives Involving Two Parties}



\subsubsection{Garble Circuit}
Garbled circuit (GC) is a cryptographic protocol \cite{yao1986how} that enables two parties (defined as the Garbler and the Evaluator) to jointly compute a function over the private data.
First, the function will be represented as a Boolean circuit.
The Garbler garbles the Boolean circuit and generates a garbled table, which is then sent to the Evaluator. 
The Evaluator receives the garbled table by the OT and evaluates the table to the result.

\subsubsection{Secret Sharing}

Secret sharing (SS) is the most basic scheme of MPC building blocks over rings.
For the commonly used 2-out-of-2 arithmetic secret sharing,
the algorithm $\texttt{Share}(\cdot)$ additively splits an $l$-bit value $x$ in the integer ring $\mathbb{Z}_{2^l}$ as the sum of two values, denoted by $\langle x \rangle_0$ and $\langle x \rangle_1$.
$x$ can be also reconstructed
as $\langle x \rangle_0 + \langle x \rangle_1 \mod 2^l$ by the algorithm $\texttt{Reconst}(\cdot, \cdot)$.
In the 2PC framework, $x$ is secretly shared with the server holding
$\langle x \rangle_0$ and client holding $\langle x \rangle_1$.
Throughout the entire inference process, intermediate values are secretly shared without any private data exposed.

\subsubsection{Oblivious Transfer}
Oblivious transfer (OT) is a fundamental cryptographic primitive that has been extensively used in different protocols, including MPC, zero-knowledge proofs (ZKP), and private set intersection (PSI).
, with a particular emphasis on PPML.
OT allows a sender to input two messages $m_0, m_1$ and a receiver to input one choice bit $b \in \{0, 1\}$ and then lets the receiver obtain the message $m_b$ based on $b$. 
For security, $b$ is kept secret against the sender and the receiver learns nothing about $m_{1-b}$. 

Most PPML protocols first generate Correlated OT (COT) correlations and then transform them to standard OTs in a blazing fast way using a correlation robust hash function (CRHF) \cite{bellare2013efficient} and the ``pre-computing OT'' technique \cite{beaver1995precomputing}. 
COT allows a sender to obtain two {\em random} strings $m_0, m_1$ where $m_0\oplus m_1=\Delta$, and makes a receiver get a {\em uniform} choice bit $b \in \{0, 1\}$ and the string $m_b=m_0\oplus b\Delta$, where $\Delta$ is fixed for multiple COT correlations.


As OT must involve public-key cryptography (PKC), it usually suffers from poor efficiency due to high computation and communication costs. 
OT extension (OTE) is hence introduced and allows two parties to extend a small number of PKC-based OT correlations to a large number of OT correlations (even any polynomial number of OTs), thus, significantly reducing computation and communication costs. Up to now, there are three types of OTE: IKNP~\cite{ishai2003extending}, PCG~\cite{boyle2018compressing}, and PCF~\cite{couteau2023pseudorandom}. 
Among them, PCG-style OTE protocols achieve the best performance.


\subsubsection{Homomorphic Encryption}

Homomorphic encryption (HE) enables computations on encrypted data such that $f(\text{Enc}(x)) = \text{Enc}(f(x))$. 
In PPML, the most widely used HE schemes are BFV~\cite{fan2012somewhat}, BGV~\cite{brakerski2014leveled}, and CKKS~\cite{cheon2017homomorphic}, all based on the RLWE problem. They support encoding 1D vectors and allow homomorphic addition, polynomial multiplication, element-wise multiplication, and rotation. Except for rotation (single-input), all operations support both ciphertext-ciphertext and ciphertext-plaintext modes.
The 1D polynomial multiplication and element-wise multiplication correspond to coefficient encoding and SIMD (Single Instruction, Multiple Data) encoding, respectively. These encoding methods will be explained in detail in Section~\ref{sssec:he-based-protocol}. Rotation $\pi_i$ shifts the encoded vector to the right by $i$ positions. For example, $\pi_1([0,1,2,3])=[3,0,1,2]$. The key differences between BFV/GV and CKKS include: (1) Data type: BFV and BGV only support integer plaintext, whereas CKKS can handle floating-point data. (2) Precision: BFV/GV yield exact results; CKKS introduces approximation errors due to residual noise after decryption. (3) Scale management: CKKS uniquely supports a rescaling operation, which allows homomorphic division by a scalar. This operation can control the scaling factor in fixed-point representations and helps mitigate bit-width growth~\cite{BLB2025}. BFV/GV do not support this operation, which is why CKKS is the preferred choice for FHE PPML pipelines.

An alternative LWE-based scheme, TFHE~\cite{chillotti2020tfhe}, supports arbitrary lookup tables via blind rotation, but encodes only a single element per ciphertext, leading to much lower efficiency than RLWE-based schemes.



\subsection{Important Features of PPML}


\subsubsection{Security and Threat Models}
One of the most critical components in a PPML framework is the threat model, which fundamentally influences both the design and the security guarantees of the PPML framework. 
Broadly, threat models can be categorized into the semi-honest threat model and the malicious threat model.

\textbf{Semi-honest.}
Most PPML studies adopt the semi-honest threat model (or honest-but-curious) \cite{rathee2020cryptflow2,rathee2021sirnn,zeng2023copriv,xu2024privquant,zeng2024eqo,jha2021deepreduce,cho2022selective,mishra2020delphi,juvekar2018gazelle,zeng2025mpcache}.
This means the participating parties are expected to follow the execution requirement of the protocols but also try to find out more information than what they are allowed to find out beyond the final result.

\textbf{Malicious.}
The malicious threat model means that participating parties can arbitrarily diverge from the normal execution of the protocol to gain some advantages.
Compared with the semi-honest threat model, there are fewer studies focusing on the malicious threat model.
For example, ABY3 \cite{mohassel2018aby3} builds the 3-party computation (3PC) protocols in the malicious threat model scenario. Falcon \cite{wagh2020falcon} adopts three servers and assumes an honest majority. FLASH \cite{byali2019flash}, SWIFT \cite{koti2021swift}, Trident \cite{chaudhari2019trident}, and Tetrad \cite{koti2021tetrad} provide a malicious threat model with at most one corruption.

\textbf{Comparison.}
The malicious threat model is more realistic and is also often more difficult than the semi-honest threat model.
In contrast, the semi-honest threat model usually leads to more efficient protocols than the malicious model and, thus, is generally used in the PPML research.
Therefore, in this survey, we mainly focus on the PPML studies built on the semi-honest threat model and organize the evolution of these studies.


\subsubsection{Interactive and Non-Interactive Protocols}

As illustrated in Figure \ref{fig:2pc_framework}, the 2PC framework can be categorized into interactive and non-interactive inference.
\textbf{Interactive inference (MPC)} requires multiple rounds of communication between the client and the server (or among multiple parties). This pattern requires significant communication due to the frequent data transfer layer by layer. Linear layers are typically computed using HE or OT, while non-linear layers are computed using GC or OT.
\textbf{Non-interactive inference (FHE)} executes all computations in a single round of communication, with the client pre-processing their inputs and the server computing using HE and returning the final results without any decryption. This pattern requires much less communication than MPC but suffers from much more computation overhead.
In Section \ref{sec:graph}, we elaborate graph-level optimizations for both protocols.

\subsection{Primary Evaluation Setups in PPML}


\subsubsection{Network Setups.}
LAN and WAN are two commonly used network setups for PPML efficiency evaluation.
\textbf{For LAN setting,} the bandwidth between the machines is 377 MBps and the echo latency is 0.3ms.
\textbf{For WAN setting,} the bandwidth between the machines is 40 MBps and the echo latency is 80ms.

\subsubsection{Evaluation Metrics.}
In the field of PPML, the studies mainly focus on the evaluation of latency and communication overhead.
\textbf{Latency} refers to the total time required to complete the inference task with the PPML protocols. LAN and WAN network settings are commonly used for evaluation.
\textbf{Communication} refers to the data exchange between different parties during the PPML protocol execution.

For example, according to early representative HE-based CryptoNets \cite{gilad2016cryptonets}, for a batch input of 4096 MNIST images, it takes 250 seconds to run the model, 44.5 seconds to execute encoding and encryption, and 3 seconds to execute decryption and decoding.
It requires around 370 MB of communication.
The classical OT-based framework CrypTFlow2 \cite{rathee2020cryptflow2} requires 3611.6 seconds (WAN) and 370.84 GB of communication to run a ResNet-50 on ImageNet.
The significant inference overhead for different PPML protocols motivates us to explore efficiency optimizations across the protocol level, model level, and system level.
\section{Protocol-Level Optimization}
\label{sec:protocol}






\subsection{Linear Layer Optimization}
\label{sec:protocol_linear}

Linear layers primarily appear as convolution layers in CNNs, attention layers and FFN layers in Transformers.
In this section, we discuss OT-based, HE-based, replicated SS (RSS)-based, and function SS (FSS)-based protocols.
Finally, we introduce the pre-processing optimization for efficient online inference.


\subsubsection{OT-Based Protocol}



We first introduce the basic workflow to leverage OT to compute the linear layers, and then introduce the OT-based protocols.

\begin{figure}
    \centering
    \includegraphics[width=\linewidth]{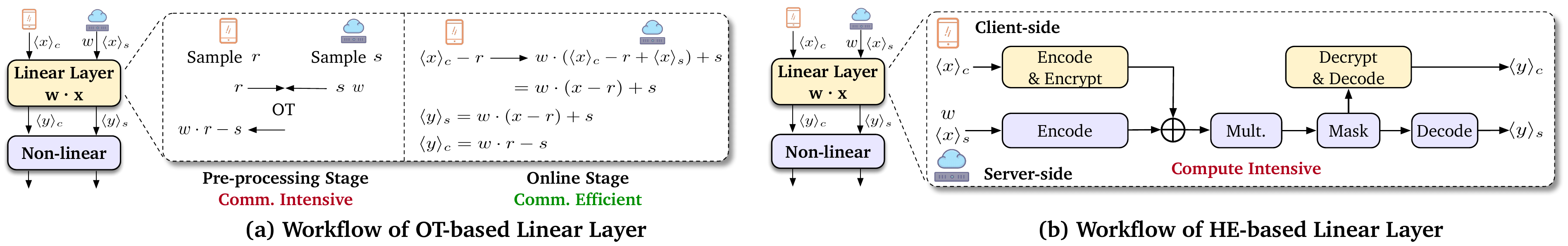}
    \caption{Basic protocol workflow of (a) OT-based linear layer computation, including a pre-processing stage and an online stage to process client's input; (b) HE-based linear layer computation.
    }
    \label{fig:matmul_workflow}
\end{figure}

\textbf{Workflow.}
As shown in Figure \ref{fig:matmul_workflow}(a), the client holds $\langle x\rangle_c$ and the server holds and $\langle x\rangle_s$ and weight $w$ where $w$ is not secretly shared in typical PPML scenarios.
To generate the result $y$ of a linear layer,
the server and client jointly execute computation in the pre-processing (offline) stage and online stage \cite{mishra2020delphi}. 
In the \textbf{pre-processing stage}, the client and server first sample random $r$ and $s$, respectively.
Then, $\langle y\rangle_c=w\cdot r-s$ can be computed with a single OT if $r \in \{0, 1\}$ for 1 bit.
With the vector optimization \cite{hussain2021coinn}, one OT can be extended to compute $w \cdot \mathbf{r} - \mathbf{s}$,
where $\mathbf{r}$ and $\mathbf{s}$ are both vectors.
When $w$ has $l_w$ bits, we can repeat the OT protocol $l_w$ times by computing $w^{(b)} \cdot r-s$ each time,
where $w^{(b)}$ denotes the $b$-th bit of $w$. The final results can then be acquired by shifting and adding the partial results together.
Compared with the pre-processing stage, the \textbf{online stage} only requires very little communication to obtain $\langle y\rangle_s=w\cdot(x-r)+s$.

\textbf{Cost analysis.}
Consider a general condition $Y = WX$, 
where $W\in \mathbb{R}^{M\times L}$, $X\in \mathbb{R}^{L\times N}$ and $Y\in \mathbb{R}^{M\times N}$.
With one round of OT, we can compute $W_{i, j}^{(b)} \cdot X_{j, :}$ for the $b$-th bit of $W_{i, j}$ and $j$-th row of $X$.
Then, the $i$-th row of $Y$, denoted as $Y_{i,:}$, can be computed as
$
    \label{eq:ot}
    Y_{i, :} = \sum_{j=0}^{L-1} \sum_{b=0}^{l_w-1} 2^b \cdot W_{i, j}^{(b)} \cdot X_{j, :},
$
where $l_w$ denotes the bit width of $W$.
Hence, to compute $Y_{i, :}$, in total $O(l_wL)$ OTs are invoked.
In each OT, the communication scales proportionally
with the vector length of $X_{j, :}$, i.e., $O(N l_x)$, where $l_x$ denotes the bit width of $X$.
The total communication of the GEMM thus becomes $O(MLN l_w l_x)$.
\textbf{Hence, the total communication of a GEMM scales with both the operands' bit widths,
i.e., $l_x$ and $l_w$, and the number of multiplications, i.e., $MLN$,}
both of which impact the round of OTs and the communication per OT.
Convolutions follow a similar pattern as GEMM.

\begin{center}
\begin{tcolorbox}[
    colback=myblue!50,
    colframe=myblue,
    width=\linewidth,
    arc=2mm, auto outer arc,
    boxrule=1pt,
    enhanced,
    drop shadow
]
    \raisebox{-0.2cm}{\includegraphics[width=0.55cm]{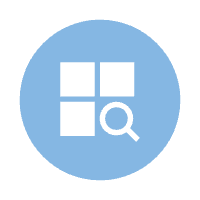}}
    \textbf{Therefore, there are two primary approaches to improve the OT-based linear layer protocol:}
    \begin{itemize}
        \item Way 1: Reduce the number of multiplications such as the Winograd convolution algorithm and the circulant convolution algorithm.
        \item Way 2: Reduce the required bit width of both activations and weights for low-precision inference. We further introduce PPML-friendly quantization algorithms in Section \ref{sec:quant}.
    \end{itemize}
\end{tcolorbox}
\end{center}

CryptFlow2 \cite{rathee2020cryptflow2} is a classic framework that proposes to use COT to compute the convolution layers, and SiRNN \cite{rathee2021sirnn} further extends the protocols to lower and mixed bit widths with more complex protocols like bit with extension and wrap.
Based on OT-based computation, CoPriv \cite{zeng2023copriv} further incorporates Winograd transformation into the convolution protocol, achieving significant multiplication and communication reduction.
CipherGPT \cite{hou2023ciphergpt} optimizes the GEMM using VOLE to significantly reduce the communication overhead.
PrivQuant \cite{xu2024privquant} proposes the protocol with low precision and EQO \cite{zeng2024eqo} further seamlessly combines Winograd convolution and quantization to achieve extreme efficiency improvement. 

    

\subsubsection{HE-Based Protocol}
\label{sssec:he-based-protocol}


The linear layer mainly includes convolutions (Conv) and matrix multiplications (MatMul), which can be efficiently evaluated by RLWE-based HE schemes with minimal communication. These HE schemes operate over polynomials with 1-dimensional coefficient vectors while DNNs compute over tensors. Thus, \textbf{Encoding} is required to map a tensor to a vector and directly determines the computational efficiency. There are two types of computation: ciphertext-ciphertext (ct-ct) and ciphertext-plaintext (ct-pt). The former is much more challenging since changing the ciphertext encoding involves heavy HE operations.
We begin by categorizing these encoding methods into three categories: \textbf{SIMD encoding, coefficient encoding, and nested encoding}. Table~\ref{tab:he_linear} provides an overview of related work. The encoding consistency refers to whether the input and output have the same encoding rule so that we can conduct consecutive Convs and MatMuls without changing encoding algorithms. Below we use $(d_1,d_2,d_3)$ to denote the dimension of a MatMul where $W\in \mathbb{R}^{d_3\times d_2}, X\in \mathbb{R}^{d_2\times d_1}$ and $Y=W\times X\in \mathbb{R}^{d_3\times d_1}$. And we use $h,w,C,K,R$ to denote the dimension of a Conv where $X\in \mathbb{R}^{h\times w\times C}, W\in \mathbb{R}^{R\times R\times C\times K}$ and $Y=W\circledast X \in \mathbb{R}^{h\times w\times K}$, where $\circledast$ denotes convolution, and we use a default zero padding to ensure the input and output have the same resolution.
\begin{table}[!tb]
\centering
\caption{HE-based linear layer protocols. 
}
\label{tab:he_linear}
\resizebox{\linewidth}{!}{
\begin{tabular}{c|c|c|c|c|l}
\toprule
\tabincell{c}{Encoding \\ Type} & Literature & Operator & \tabincell{c}{Encoding \\ Consistency} &Data Type  &Technique \\
    \midrule
    \multirow{6}{*}{SIMD} &Gazelle~\cite{juvekar2018gazelle} & Conv, MatMul & \checkmark$^*$ & ct-pt &   Baseline protocol, Hoisting optimization  \\
    & HEAR~\cite{kim2022secure}, MPCNN \cite{lee2022low} & Conv, MatMul & \checkmark$^*$ & ct-pt &   Multiplexed format        \\
    &Delphi~\cite{mishra2020delphi} & Conv, MatMul & \checkmark& ct-pt &   Pre-processing optimization        \\
    &BOLT~\cite{pang2024bolt} & MatMul & \checkmark& ct-pt&   BSGS optimization, reduce rotations        \\
    \cmidrule{2-6}
    & E2DM~\cite{jiang2018secure} & MatMul & \checkmark & ct-ct & Multiplication of two ciphertext square matrices \\
    &BOLT~\cite{pang2024bolt} & MatMul & \textcolor{black}{\ding{55}} & ct-ct&  low multiplication depth ct-ct MatMul for Transformer   \\
    & Powerformer~\cite{zimerman2024power} & MatMul & \textcolor{black}{\ding{55}}& ct-ct & Consecutive MatMuls in Transformer \\
    & BLB~\cite{BLB2025} & MatMul & \textcolor{black}{\ding{55}}& ct-ct & Consecutive MatMuls in Transformer \\
    \midrule
    \multirow{6}{*}{Coefficient} & Cheetah~\cite{huang2022cheetah} & Conv, MatMul & \textcolor{black}{\ding{55}}& ct-pt &  The first coefficient encoding protocols      \\
    &Iron~\cite{hao2022iron} & MatMul & \textcolor{black}{\ding{55}}& ct-pt & Optimized encoding for MatMul    \\
    &Falcon~\cite{xu2023falcon} & Conv & \textcolor{black}{\ding{55}}& ct-pt & Optimized encoding for MatMul depthwise convolution \\
    &ConvFHE~\cite{kim2023optimized} & Conv & \checkmark & ct-pt &   Output repacking for batch Conv        \\
    &Bumblebee~\cite{lu2023bumblebee} & MatMul & \textcolor{black}{\ding{55}}& ct-pt &   Output repacking to reduce communication       \\
    &Rhombus~\cite{he2024rhombus} & MatMul & \textcolor{black}{\ding{55}}& ct-pt &   Efficient output repacking\\
    \midrule
    \multirow{2}{*}{Nested}
    & NeuJeans~\cite{ju2024neujeans} & Conv & \checkmark & ct-pt & Propose nested encoding and bootstrapping-based fusion \\
    & PrivCirNet~\cite{xu2024privcirnet} & Conv, MatMul & \checkmark & ct-pt & Nested encoding for block circulant Conv and MatMul \\
    \bottomrule
    \multicolumn{4}{l}{$*$ Encoding is unconsistent for Conv with stride > 1.}
    \end{tabular}
    }
\end{table}

\begin{table}[!tb]
    \renewcommand{\arraystretch}{1.0}\
    \Huge
    \centering
    \caption{Theoretical computation complexity of different works, where $n$ is the number of elements a ciphertext can hold.}
    \label{tab:complexity_compare}
    \resizebox{\linewidth}{!}{
    \begin{tabular}{c|ccc|ccc}
    \toprule 
    \multirow{2}{*}{Framework} &  \multicolumn{3}{c|}{MatMul} & \multicolumn{3}{c}{Conv} \\
    \cmidrule{2-7}
    &  \multicolumn{1}{c}{\# HE-Pmult}&\# HE-Rot & \# Ciphertexts & \multicolumn{1}{c}{\# HE-Pmult}&\# HE-Rot & \# Ciphertexts\\
    \midrule
    Gazelle~\cite{juvekar2018gazelle} & $O(d_1d_2d_3/n)$ &$O(d_1(d_2+d_3)/n+d_3)$ & $O(d_1(d_2+d_3)/n)$ & $O(hwCK/n)$ & $O(hw(C+K)/n+K)$ & $O(hw(C+K)/n)$\\
    \midrule
    HEAR~\cite{kim2022secure}, MPCNN\cite{lee2022low}&$O(d_1d_2d_3/n)$&$O(d_1d_2d_3/n)$&$O(d_1(d_2+d_3)/n)$   & $O({hwKCR^2}/{n})$      & $O({hwKCR}/{n})$        & $O({hw(K+C)}/{n})$ \\
    \midrule
    \multirow{1}{*}{Cheetah~\cite{huang2022cheetah}}& $O(d_1d_2d_3/n)$ &\multirow{1}{*}{0} &$O(d_1d_2/n+\lceil d_1/n\rceil d_3)$ &$O(hwCK/n)$ & \multirow{1}{*}{0} & $O(hwC/n+\lceil hw/n\rceil K)$  \\
    \midrule
    \multirow{1}{*}{Iron~\cite{hao2022iron}}& $O(d_1d_2d_3/n)$ &\multirow{1}{*}{0} &$O(\sqrt{d_1d_2d_3/n})$ & $O(hwCKR^2/n)$ &\multirow{1}{*}{0} & $O(\sqrt{hwCKR^2/n})$ \\
    \midrule
    \multirow{1}{*}{Bumblebee~\cite{lu2023bumblebee}}& $O(d_1d_2d_3/n)$ &$O(d_1d_3\log_2n/(2\sqrt{n}))$ & $O(d_1(d_2+d_3)/n)$ &$O(hwCK/n)$ & $O(hwK\log_2n/(2\sqrt{n}))$ & $O(hw(C+K)/n)$\\
    \midrule
    \multirow{1}{*}{Neujeans~\cite{ju2024neujeans}}& $O(d_1d_2d_3/n)$ &$O(\sqrt{d_1d_2d_3/n})$  &$O(d_1(d_2+d_3)/n)$ &$O(hwCK/n)$ & $O(\sqrt{hwCK/n})$ & $O(hw(C+K)/n)$ \\
    \midrule
    \multirow{1}{*}{Bolt~\cite{pang2024bolt}}& $O(d_1d_2d_3/n)$ &$O(\sqrt{d_1d_2d_3/n})$  &$O(d_1(d_2+d_3)/n)$ &$O(hwCKR^2/n)$ & $O(\sqrt{hwCKR^2/n})$ & $O(hw(CR^2+K)/n)$ \\
    \midrule
\multirow{2}{*}{PrivCirNet\cite{xu2024privcirnet}}&\multicolumn{3}{c|}{MatMul with circulant weight matrix} & \multicolumn{3}{c}{Conv with circulant weight kernel}\\
    \cmidrule{2-7}
    & $O(d_1d_2d_3/(nb))$ &$O(\sqrt{d_1d_2d_3/(nb)})$  &$O(d_1(d_2+d_3)/n)$&$O(hwCK/(nb))$ & $O(\sqrt{hwCK/(nb)})$ & $O(hw(C+K)/n)$\\
    \bottomrule
    \end{tabular}
    }
\end{table}

\begin{figure}[!tb]
    \centering
    \includegraphics[width=\linewidth]{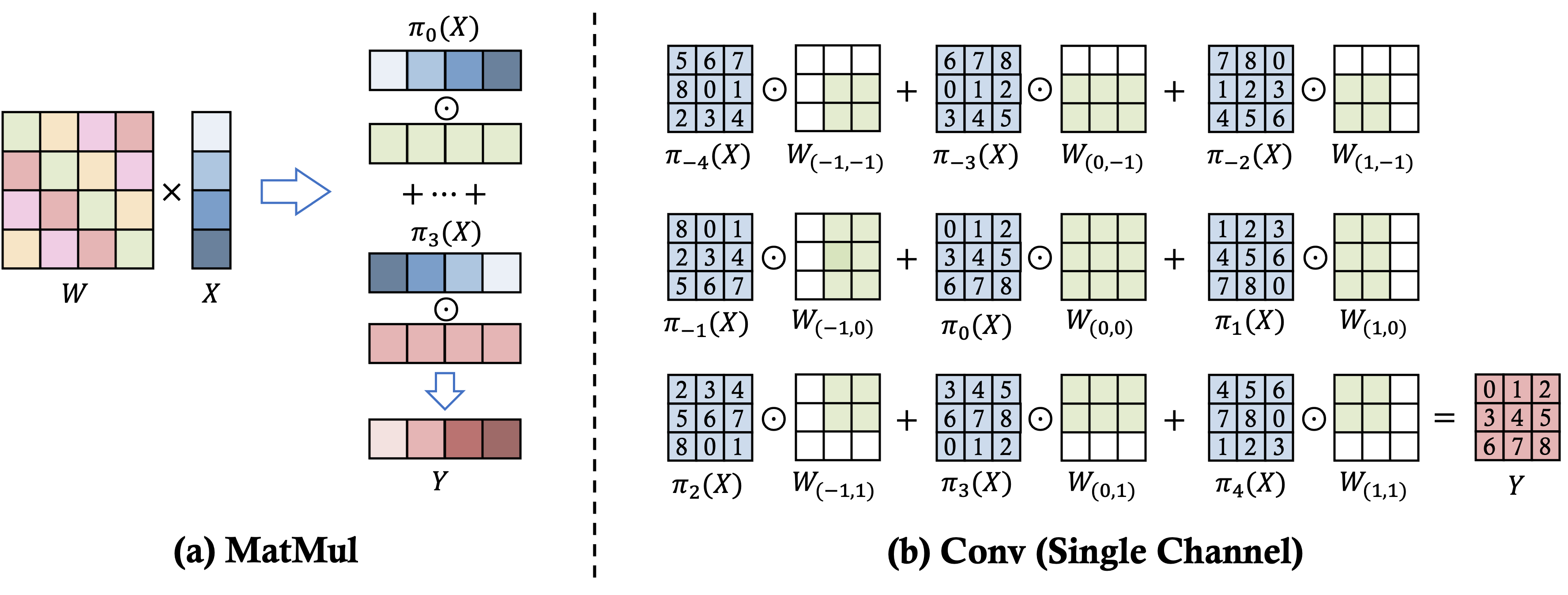}
    \caption{Toy example of (a) MatMul protocol and (b) Conv protocol using SIMD encoding in Gazelle \cite{juvekar2018gazelle}.}
    \label{fig:gazelle}
\end{figure}

\textbf{SIMD encoding.} We begin with SIMD encoding, the first proposed scheme~\cite{gilad2016cryptonets} that supports homomorphic element-wise addition and multiplication. Gazelle is a representative work that extends SIMD encoding to both MatMul and Conv. As shown in Figure~\ref{fig:gazelle}, Gazelle uses diagonal encoding for MatMul, where each diagonal of the weight matrix is encoded into a polynomial, and input rotations $\pi_i(X)$ are applied to align slots. For Conv, multiple rotations and ciphertext-plaintext multiplications (PMult) are performed using different weight polynomials. Gazelle requires $O(d_1d_2d_3/n)$ rotations and PMults for MatMul, and $O(hwCKR^2/n)$ for Conv. It also introduces a hoisting optimization to reduce redundant computation in repeated rotations. Since HE rotations are significantly more expensive than ct-pt multiplications, many subsequent works focus on reducing the number of rotations.
Delphi proposes a pre-processing technique where multiplication triples $\langle A\rangle, \langle B\rangle, \langle C\rangle$ with $C = A \times B$ are generated offline as secret shares. This shifts all linear homomorphic computations to the offline phase, requiring only plaintext linear operations online. However, it is only suitable for HE+MPC hybrid frameworks due to the reliance on secret sharing.
HEAR~\cite{kim2022secure} introduces a \textit{multiplexed format}, simplifying the rearrangement of convolution outputs, especially when the stride is larger than one, though its basic complexity remains similar to Gazelle. MPCNN~\cite{lee2022low} leverages the multiplexed format to repack downsampled outputs into a more compact form, improving convolution efficiency.

BOLT~\cite{pang2024bolt} improves Gazelle’s MatMul protocol by introducing column-packing, which prioritizes the $d_3$ dimension to reduce slot-wise summation. It further applies Baby-Step Giant-Step (BSGS) optimization to minimize rotation overhead. Theoretically, BOLT achieves $O(\sqrt{d_1 d_2 d_3 / n})$ rotations and $O(d_1 d_2 d_3 / n)$ PMults, with ciphertext I/O reduced to $O(d_1(d_2 + d_3)/n)$—the current state-of-the-art for ct-pt MatMul.

The above schemes focus on ct-pt operations. For ct-ct settings, encoding becomes more involved. E2DM~\cite{jiang2018secure} proposes a ct-ct MatMul protocol for square ciphertext matrices, requiring multiplicative depth 3, with $O(d)$ rotations and $O(d)$ ct-ct multiplications. For neural networks like Transformers that involve ct-ct MatMul, BOLT introduces an alternative encoding that reduces the multiplicative depth to 1 at the cost of increasing rotations to $O(d \log d)$, improving overall efficiency since HE cost scales with depth.
PowerFormer~\cite{park2024powerformer} improves E2DM by optimizing ct-ct MatMul for non-square matrices using replication to reshape them into square form, reducing rotation overhead. However, this method is limited to cases where the matrix aspect ratio is an integer.
Building on BOLT~\cite{pang2024bolt}, BLB~\cite{BLB2025} introduces further optimizations. It reduces the number of accumulations within a ciphertext via pre-processing and introduces multi-head parallel packing. Encoding different Transformer heads into one ciphertext significantly reduces the rotation cost. Compared to BOLT and PowerFormer, BLB reduces the number of rotations by $29\times$ and $8\times$, respectively. Although it has a multiplicative depth of 3, the drastically reduced rotation count leads to significantly better overall performance.
\begin{center}
\begin{tcolorbox}[
    colback=myblue!50,
    colframe=myblue,
    width=\linewidth,
    arc=2mm, auto outer arc,
    boxrule=1pt,
    enhanced,
    drop shadow
]
    \raisebox{-0.2cm}{\includegraphics[width=0.55cm]{icon/classify.png}}
    \textbf{Improvements to SIMD Encoding focus on reducing rotations, with two main approaches:}
    \begin{itemize}
        \item Way 1: Packing as few elements as possible that need to be accumulated in a ciphertext, e.g., column packing in BOLT~\cite{pang2024bolt}.
        \item Way 2: Reuse rotation results as much as possible, e.g., BSGS optimization.
    \end{itemize}
\end{tcolorbox}
\end{center}


\begin{figure}[!tb]
    \centering
    \includegraphics[width=\linewidth]{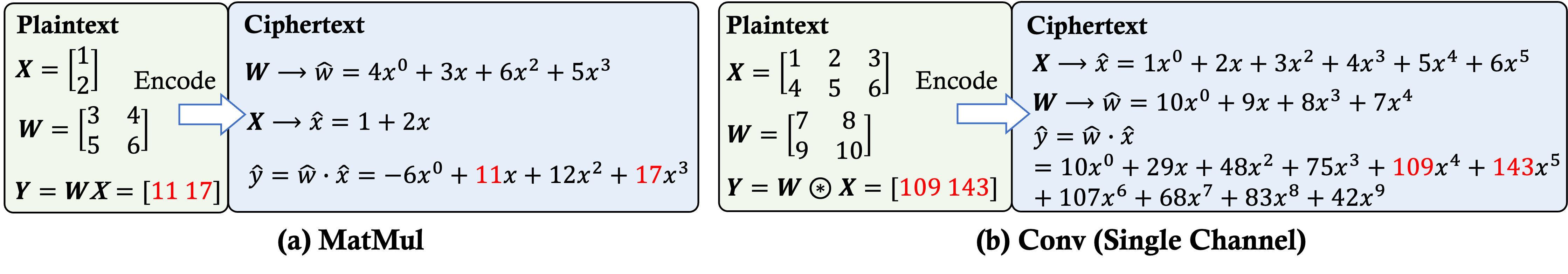}
    \caption{Toy example of (a) MatMul protocol and (b) Conv protocol using coefficient encoding.
    }
    \label{fig:coeff}
\end{figure}

\textbf{Coefficient encoding.}
In HE+MPC hybrid frameworks, OT-based protocols are more efficient over the ring $\mathbb{Z}_{2^l}$ than the field $\mathbb{Z}_p$, and modulo reduction over $\mathbb{Z}_{2^l}$ is more hardware-friendly. However, traditional homomorphic SIMD encoding requires computation over $\mathbb{Z}_p$. To better support non-linear protocols on $\mathbb{Z}_{2^l}$, Cheetah~\cite{huang2022cheetah} implements Conv and MatMul using coefficient encoding by leveraging the similarity between polynomial multiplication and convolution.

Figure~\ref{fig:coeff} shows toy examples of Cheetah's MatMul and Conv. For MatMul, the input $X$ is split into vectors and encoded as polynomial coefficients, while the weights $W$ are encoded in reverse order. For Conv, $X$ is encoded in row-major order, and the kernel weights are encoded in reverse, with coefficients from different rows spaced by $W$. Only part of the resulting ciphertext is useful and is extracted as LWE ciphertexts for subsequent 2PC operations. Cheetah achieves $O(d_1d_2d_3/n)$ PMults for MatMul and $O(hwCK/n)$ PMults for Conv, without requiring expensive rotations.
Building on Cheetah, Iron~\cite{hao2022iron} introduces an optimization for MatMul by partitioning the $(d_1,d_2,d_3)$ MatMul into smaller blocks of size $(d_1',d_2',d_3')$ such that $d_1'd_2'd_3' \leq n$. This ensures correct computation within each block and reduces transmitted ciphertexts to $\sqrt{\frac{2d_1d_2d_3}{n}}$.
For visual models like MobileNetV2 and EfficientNet, Falcon~\cite{xu2023falcon} optimizes depthwise convolutions by zero-aware greedy packing and communication-aware tiling, reducing overhead caused by padded zero channels.

To improve the packing density, 
ConvFHE~\cite{kim2023optimized} and BumbleBee~\cite{lu2023bumblebee} support packing $C$ RLWE ciphertexts into a single RLWE ciphertext, called \textsf{PackRLWEs} in Rhombus~\cite{he2024rhombus}. This requires $O(C)$ homomorphic automorphisms (\textsf{HomAuto}, or slot rotations). Using \textsf{PackRLWEs}, ConvFHE proposes a new encoding that packs multiple convolution outputs into one ciphertext, reducing I/O overhead.
BumbleBee improves \textsf{PackLWEs} by reducing \textsf{HomAuto} overhead, and applies ciphertext interleaving for MatMul, arranging outputs densely via \textsf{PackRLWEs}, reducing transmission to $O(d_1(d_2+d_3)/n)$—matching SIMD encoding.
Rhombus~\cite{he2024rhombus} further reduces \textsf{PackRLWEs} complexity from $O(C)$ to $O(\sqrt{C})$ using a split-point selection technique. It also proposes dimension-aware strategies, choosing between row-major and column-major Conv layouts.

\begin{center}
\begin{tcolorbox}[
    colback=myblue!50,
    colframe=myblue,
    width=\linewidth,
    arc=2mm, auto outer arc,
    boxrule=1pt,
    enhanced,
    drop shadow
]
    \raisebox{-0.2cm}{\includegraphics[width=0.55cm]{icon/classify.png}}
    \textbf{Due to the sparsity of results in ciphertext distribution under coefficient encoding, improvements to coefficient encoding focus on efficiently reducing the number of I/O ciphertexts transmitted. There are two primary approaches to achieve this:}
    \begin{itemize}
        \item Way 1: Optimizing the dimensional design for block-wise matrix computation, e.g., Iron \cite{hao2022iron} and Rhombus \cite{he2024rhombus}.
        \item Way 2: Packing valid coefficients from different ciphertexts into a single ciphertext using \textsf{HomAuto}, while minimizing the number of \textsf{HomAuto} operations, e.g., ConvFHE \cite{kim2023optimized} and BumbleBee \cite{lu2023bumblebee}.
    \end{itemize}
\end{tcolorbox}
\end{center}

\textbf{Nested encoding.}
Neujeans~\cite{ju2024neujeans} finds that SIMD encoding and Coefficient encoding can be transformed to each other through the discrete Fourier transform (DFT) as shown in Lemma~\ref{lemma:DFT}. The main reason is that polynomial multiplication in the coefficient domain is equivalent to element-wise multiplications in the frequency domain, leading to Lemma~\ref{lemma:DFT}. Based on this, Neujeans implements a nested encoding method for Conv, utilizing coefficient encoding for convolutions of each input channel and SIMD encoding for all input and output channels. This encoding scheme reduces the computational cost to $O(\sqrt{hwCK/n})$ rotations and $O(hwCK/n)$ PMults, while preserving the same input/output consistency, making it the current best encoding algorithm for Conv. PrivCirNet~\cite{xu2024privcirnet} further extends the nested encoding for block circulant Conv and MatMul, reducing the number of rotations to $O(\sqrt{hwCK/nb})$ and the number of PMults to $O(hwCK/nb)$. nested encoding makes use of the advantages of both SIMD and coefficient encoding. However, nested encoding requires DFT to input, which incurs huge computation costs when the input is in the ciphertext domain. To address this, Neujeans proposes fusing convolution with bootstrapping, as bootstrapping inherently involves computing the DFT, allowing the DFT overhead to be amortized and thereby avoiding additional computational cost.
\begin{lemma}
    \label{lemma:DFT}
    $\lj \operatorname{DFT}(w) \rj_{\mathrm{SIMD}} \times \lj \operatorname{DFT}(x) \rj_{\mathrm{SIMD}} = \operatorname{DFT}(\lj w \rj_{\mathrm{Coeff}}\times \lj x \rj_{\mathrm{Coeff}})$
\end{lemma}

\textbf{Theoretical complexity.} 
We summarize the computational and communication complexities of existing encoding schemes for ct-pt Conv and MatMul in Table~\ref{tab:complexity_compare}. For convolution, the nested encoding scheme achieves the best overall performance, while for matrix multiplication, both BOLT and nested encoding offer optimal efficiency.
On the other hand, coefficient encoding is more computation-efficient; however, it suffers from input-output layout mismatches, making it unsuitable for consecutive linear operations and limiting its applicability in practical scenarios.
\begin{center}
\begin{tcolorbox}[
    colback=orange!8,
    colframe=orange!20,
    width=\linewidth,
    arc=2mm, auto outer arc,
    boxrule=1pt,
    enhanced,
    drop shadow
]
    \raisebox{-0.2cm}{\includegraphics[width=0.5cm]{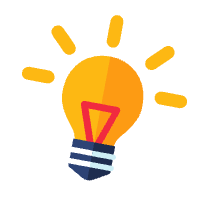}}
    \textbf{Takeaways of different encoding methods:}
    \begin{itemize}
        \item Most SIMD encoding can maintain the consistency of input and output encoding, allowing consecutive computation, and is suitable for FHE scenarios.
        \item Coefficient encoding eliminates rotation and has lower computational complexity. It has significantly lower computational/communication complexity for convolution than SIMD, but the encoding is inconsistent and is only used for single-layer computation in HE/MPC hybrid protocol scenarios.
        \item Nested encoding is suitable for parallel processing of multiple convolution scenarios. A single convolution can take advantage of coefficient encoding, and parallel convolution processing can take advantage of SIMD. It is suitable for consecutive computing scenarios, but consecutive calculation requires DFT of the ciphertext, which needs to be integrated with Bootstrapping to reduce overhead.
    \end{itemize}
\end{tcolorbox}
\end{center}


\subsubsection{RSS-Based and FSS-Based Protocol}
In this section, we introduce replicated SS (RSS)-based protocols and function secret sharing (FSS)-based protocols.

\textbf{RSS.}
3PC protocols like PUMA \cite{dong2023puma} and ABY3 \cite{mohassel2018aby3} are built based on the 2-out-of-3 replicated secret sharing (RSS), where
a secret value $x\in 2^\ell$ is shared by three random values $x_0, x_1, x_2 \in 2^\ell$ such that $x=x_0+x_1+x_2 \pmod{2^\ell}$, and party $P_i$ gets $(x_i, x_{i+1})$ (denoted as $[\![x]\!]$). 

Let $(c_1$, $c_2$, $c_3)$ be public constants and $([\![x]\!], [\![y]\!])$ be two secret-shared values. The secure addition and multiplication procedures are as follows:
\begin{itemize}
    \item \textit{Addition.} $[\![c_1 x+c_2 y+c_3]\!]$ can be computed as $(c_1 x_0+c_2 y_0+c_3, c_1 x_1+c_2 y_1, c_1 x_2+c_2 y_2)$, where $P_i$ can compute its share locally. When $(c_1 =1,c_2 =1,c_3 =0)$, we get $[\![x+y]\!]$.  
    \item \textit{Multiplication.} Parties follow steps: 
    \romannumeral1) first, $P_i$ computes $z_i = x_iy_i + x_{i+1}y_i + x_iy_{i+1}$ locally; 
    \romannumeral2) parties then perform \textit{re-sharing} by letting $P_i$ sends $z_i'=\alpha_i+z_i$ to $P_{i-1}$, where $\alpha_0+\alpha_1+\alpha_2=0$ ($P_i$ can generate $\alpha_i$ using pseudorandom generators with negligible overhead as~\citet{mohassel2018aby3});
    \romannumeral3) finally, $\{(z_0', z_1'), (z_1', z_2'), (z_2',z_0')\}$ form the 2-out-of-3 replicated secret shares of $[\![xy]\!]$. 
\end{itemize}

For more details about the 3PC protocol, please refer to \citet{mohassel2018aby3,ma2023secretflow,dong2023puma}.

\textbf{FSS.}
FSS has become a trend for private inference where a trusted dealer provides input-independent correlated randomness to the two parties involved in the secure computation, leading to low online costs for 2PC inference.
LLAMA \cite{gupta2022llama} proposes FSS-based linear layers such as element-wise multiplication, matrix multiplications, and convolutions. 
Afterward, Orca \cite{jawalkar2024orca} and SIGMA \cite{gupta2023sigma} follow LLAMA to construct the FSS-based linear layers.

\subsubsection{Pre-Processing Optimization}

Most of the recent PPML works have adopted the method that divides the inference process into two phases, i.e., a pre-processing phase (offline) and an online phase to transfer the most of the online cost to the offline phase and significantly reduce the online cost \cite{mishra2020delphi,demmler2015aby}.

During the pre-processing phase, the core step is to generate input-independent Beaver's triplets, which can be directly consumed by the online phase with much lower communication.
If there are only two parties involved in the computation, Beaver's triplets can be generated by two well-known approaches: 1) OT-based generation with lower computation and 2) HE-based generation with lower communication.
For instance, MiniONN \cite{liu2017oblivious} and Delphi \cite{mishra2020delphi} use the HE-based method to generate Beaver's triplets, CoPriv \cite{zeng2023copriv} uses the OT-based method to generate Beaver's triplets while ABY \cite{demmler2015aby} and SecureML \cite{mohassel2017secureml} provide both alternatives for Beaver's triplet generation.
It is worth noting that, as mentioned in CoPriv, the total communication is dominated by the pre-processing phase and cannot be ignored in the PPML inference system.


\subsection{Non-Linear Layer Optimization}
\label{sec:protocol_nonlinear}



Non-linear layers primarily appear as ReLU in CNNs, Softmax in attention layers and GeLU/SiLU in FFN layers in Transformers.
In this section, we introduce GC-based, OT-based, FSS-based, and HE-based protocols.

\subsubsection{SS-Based Protocol}
SS-based protocols mainly include GC-based, OT-based, and FSS-based methods.

\textbf{GC-based protocols.}
GC is widely used in early PPML research for non-linear layer evaluation.
The key feature of GC is that it only requires a constant round to evaluate the function, but suffers from high communication overhead.
ABY \cite{demmler2015aby} combines arithmetic sharing, boolean sharing, and GC in one PPML framework.
MiniONN \cite{liu2017oblivious} uses GC to evaluate ReLU, max pooling, and truncation.
Delphi \cite{mishra2020delphi} uses GC to evaluate ReLU and a series of works follow Delphi to optimize the model \cite{jha2021deepreduce,cho2021sphynx,ghodsi2021circa,cho2022selective,park2022aespa,liseesaw}.
XONN \cite{riazi2019xonn} achieves an end-to-end framework based on GC with a combination of binary neural networks.
COINN \cite{hussain2021coinn} and ABNN2 \cite{shen2022abnn2} choose to evaluate non-linear layers using GC.

\textbf{OT-based protocols.}
OT serves as a fundamental building block in cryptographic primitives, capable of realizing any operation in PPML without incurring any approximation errors.
All non-linear functions can be implemented based on the basic protocols below:
\begin{itemize}
    \item Millionaires'/Wrap: The $\ell$-bit Millionaires’ functionality, $\mathcal{F}_{\mathrm{Mill}}^{\ell}$ take $x\in {\{ 0,1 \}}^{\ell}$ from $P_0$ and $y\in {\{ 0,1 \}}^{\ell}$ from $P_1$ as input and returns $\langle z\rangle^B$ such that $z=\mathbf{1}\{x<y\}$. The $\ell$-bit wrap functionality,  $\mathcal{F}_{\mathrm{Wrap}}^{\ell}$ takes the same inputs returns $\langle z\rangle^B$ such that $z=\mathbf{1}\{x<L-1-y\}$.
    \item AND: The functionality $\mathcal{F}_{\mathrm{AND}}$ takes the boolean shares $(\langle x\rangle^B,\langle y\rangle^B)$ as input and returns $\langle x\wedge y\rangle^B$. $\mathcal{F}_{\mathrm{AND}}$ can be realized using bit-triples \cite{rathee2020cryptflow2}, and the parties can generate two bit-triples through an instance of $\binom{16}{1}\mathrm{-OT}_{2}$.
    \item Boolean to Arithmetic (B2A): The $\ell$-bit B2A functionality, $\mathcal{F}_{\mathrm{B2A}}^{\ell}$, takes boolean shares $\langle x\rangle^B$ and ruturns arithmetic shares of the same value $\langle x\rangle^{\ell}$ by an instance of $\binom{2}{1}\mathrm{-OT}_{\ell}$.
    \item Multiplexer (MUX): The $\ell$-bit MUX functionality, $\mathcal{F}_{\mathrm{B2A}}^{\ell}$, takes boolean shares $\langle y\rangle^{\ell}$ and arithmetic shares and returns $\langle z\rangle^{\ell}$ such that $z=y$ if $x=0$ else $z=0$, which requires two instances of $\binom{2}{1}\mathrm{-OT}_{\ell}$.
    \item Lookup Table (LUT): For table $T$ with $M$ $n$-bits entries, The LUT functionality $\mathcal{F}_{\mathrm{LUT}}^{M,n,T}$ takes $\langle x\rangle^{m}$ as input and returns $\langle z\rangle^{n}$ such that $z=T[x]$, which requires an instance of $\binom{M}{1}\mathrm{-OT}_{n}$. 
\end{itemize}
Through the aforementioned basic protocol, we can implement various non-linear functions such as ReLU \cite{rathee2020cryptflow2,rathee2021sirnn,huang2022cheetah}, GELU \cite{hao2022iron,lu2023bumblebee,pang2024bolt}, and Softmax \cite{hao2022iron,lu2023bumblebee,pang2024bolt}.
Furthermore, truncation and extension can also be achieved \cite{huang2022cheetah,rathee2021sirnn,xu2024privquant,zeng2024eqo}.

\textbf{FSS-based protocols.}
FSS is also widely used in non-linear layers.
LLAMA \cite{gupta2022llama} applies FFS-based protocol to various non-linear math functions, including Sigmoid, tanh, and reciprocal square root with precise low bit width representations.
Orca \cite{jawalkar2024orca} proposes FSS-based protocols for CNNs with GPU acceleration. Orca builds basic building blocks, including select and sign extension, and proposes non-linear protocols, including ReLU and Softmax.
Different from Orca, SIGMA \cite{gupta2023sigma} focuses on Transformers and builds efficient FSS-based building blocks, including truncation, DReLU, and comparison. Based on the building blocks, SIGMA combines them to build complex non-linear layers, including GeLU, Softmax, and layer normalization.

\subsubsection{HE-Based Protocol}
We introduce two ways to achieve HE-based non-linear protocols, i.e., polynomial approximation and TFHE.

\textbf{Polynomial approximation.}
Evaluating non-linear layers using word-wise HE schemes (e.g., BFV, BGV, CKKS) typically relies on polynomial approximation. Low-degree polynomials, pruning techniques, or simple functional replacements can improve inference efficiency but often require model retraining, which we discuss in model-level optimizations in Section~\ref{sec:model}. When retraining is not feasible, higher-degree polynomials or indirect approximation methods are necessary to meet accuracy requirements---albeit at the cost of increased inference overhead. This section focuses on non-linear function approximation within a pure HE context.

In general, any continuous function can be approximated by polynomial interpolation and evaluated using homomorphic additions and multiplications. For numerical stability, Chebyshev bases are often preferred over power bases. However, discontinuous functions (e.g., sign functions, modular reductions) cannot be approximated directly and instead require indirect methods. The sign function is essential for comparison-based operations such as ReLU, MaxPool, Softmax, and Argmax. An analytic approximation of the sign function using the Fourier series was proposed in~\cite{boura2018chimera}, while~\citet{chialva2018conditionals} introduced a method based on an approximate $\tanh$ function. More recently, minimax polynomial approximations have been widely adopted due to their lower multiplicative depth and reduced use of non-scalar multiplications~\cite{lee2021minimax,lee2022optimization}. Nevertheless, even with such techniques, approximating ReLU with an $\ell_1$-norm error below $2^{-13}$ in an FHE-based ResNet still requires a multiplicative depth of 14~\cite{lee2022low}.


\textbf{TFHE-based protocols.}
FHE over the Torus (TFHE) was introduced in \cite{chillotti2020tfhe}, offering significantly faster sub-second bootstrapping. In TFHE, bootstrapping is programmable, enabling the evaluation of any univariate functions in a look-up table (LUT) format, making it well-suited for nonlinear layer computations. The protocol for ReLU activation was designed in \cite{boura2019simulating}, and \cite{chillotti2021programmable} extended this to MaxPool operations with precision ranging from 8 to 12 bits. Programmable bootstrapping (PBS) in TFHE incurs a computational cost of $O(2^l)$, where $l$ represents the bit width of the input. To address this, \cite{stoian2023zama_tfhe_quant} and \cite{folkerts2021redsec} applied the TFHE scheme to quantized neural networks with operands of lower bit width. However, TFHE schemes do not support SIMD operations and suffer from higher latency compared to BFV/CKKS. State-of-the-art TFHE scheme~\cite{zama2024TFHE} requires 489ms to evaluate a nonlinear function with an 8-bit input, making it challenging to apply in neural network applications due to the high latency.
Recent works, such as~\cite{liu2023amortized,liu2025relaxed,bae2024bootstrapping,bae2025bootstrapping}, aim to achieve batch PBS by combining TFHE with BFV and CKKS schemes.

\begin{center}
\begin{tcolorbox}[
    colback=orange!8,
    colframe=orange!20,
    width=\linewidth,
    arc=2mm, auto outer arc,
    boxrule=1pt,
    enhanced,
    drop shadow
]
    \raisebox{-0.2cm}{\includegraphics[width=0.5cm]{icon/light.png}}
    \textbf{Takeaways of non-linear layer protocols:}
    \begin{itemize}
        \item Existing PPML studies use GC-based, OT-based, FSS-based, and HE-based methods to evaluate non-linear functions, and they are tailored to specific application scenarios. 
        \item GC-based protocols only require a constant round but suffer from high communication costs, limiting the adoption in recent works.
        \item OT-based protocols are general for different non-linear functions but the communication cost is high due to the frequent interactive computation. VOLE can help reduce the communication costs.
        \item FSS-based protocols provide sublinear communication for specific functions such as equality or comparison, but lack generality.
        \item HE-based protocols enable low communication costs, but they face high computational overhead when evaluating high-order polynomial approximations and limited support for complex non-linear functions. TFHE leverages PBS to evaluate arbitrary non-linear functions without approximation.
    \end{itemize}
\end{tcolorbox}
\end{center}

\subsection{Graph-Level Technique}
\label{sec:graph}
Despite linear and non-linear layers, graph-level optimizations are also crucial, focusing on inter-layer techniques for interactive and non-interactive inference.

\subsubsection{Interactive Protocols with SS-HE Conversion}

Interactive protocols often combine HE with secret sharing, requiring conversion protocols~\cite{hao2022iron,huang2022cheetah}. A typical HE-to-SS conversion works as follows: the server holding $\text{Enc}(x)$ samples a random vector $r$, computes $\text{Enc}(x-r)$, and sends it to the client. The client decrypts to get $x-r$, while the server keeps $r$, forming additive shares. The reverse conversion is straightforward: the client encrypts $x-r$ and sends $\text{Enc}(x-r)$ to the server, which adds $r$ homomorphically to recover $\text{Enc}(x)$.

Most protocols follow this pattern. However, BLB~\cite{BLB2025} recently discovered that applying this to CKKS ciphertexts can lead to security issues~\cite{boemer2020mp2ml}, due to incorrect treatment of SIMD encoding. They proposed a secure alternative that performs conversion before encoding.

\subsubsection{Non-Interactive Protocol}
Non-interactive protocols mainly include bootstrapping, level consumption reduction, and lazy rescaling.

\textbf{Bootstrapping.}
Unless retraining a HE-friendly network, bootstrapping becomes unavoidable in practical neural networks due to the multiplication depth significantly exceeding the leveled HE parameter configuration. Bootstrapping for CKKS was first introduced in~\cite{cheon2018bootstrapping} to reset the modulus chain (i.e., the ciphertext level), thereby enabling continued homomorphic computation. Subsequent works~\cite{han2020better,bossuat2021efficient,bossuat2022bootstrapping} proposed optimizations targeting bootstrapping itself. 

CKKS bootstrapping comprises several sub-procedures, including \textit{coefficient-to-slots} (\textsf{CtS}), \textit{slots-to-coefficient} (\textsf{StC}), \textit{modular raising}, and \textit{approximate modular reduction}. From a graph-level optimization perspective, these steps may be fused with upstream or downstream components in the neural network computation graph to hide certain computational overhead or avoid redundant domain conversions. For example, ConvFHE~\cite{kim2023optimized} leverages domain conversion (i.e., \textsf{CtS}) within bootstrapping to integrate coefficient encoding-based $\textsf{Conv}$ into an end-to-end HE-based protocol, transitioning to the slot domain for element-wise approximate non-linear function evaluation. Similarly, NeuJeans~\cite{ju2024neujeans} fuses \textsf{Conv} with StC within bootstrapping to conceal convolution overhead, enabling more efficient CNN inference.

\textbf{Level consumption reduction via folding.}
To reduce multiplicative depth, nGraph-HE~\cite{boemer2019ngraph} introduces a series of plaintext-ciphertext multiplication folding techniques that fuse scaling operations into adjacent linear layers. These include \textit{AvgPool folding}, which replaces \textsf{AvgPool}–\textsf{Conv} sequences with scaled convolutions; \textit{activation folding}, which transforms polynomial activations into monic forms and propagates scaling factors backward; and \textit{BatchNorm folding}, which absorbs affine normalization parameters into the preceding linear layers. Similarly, MPCNN~\cite{lee2022low} applies folding in a fully homomorphic ResNet-20 by adapting batch normalization to eliminate the scaling overhead required to fit inputs into \([-1,1]\) before bootstrapping, thereby avoiding additional multiplicative depth.

\textbf{Lazy rescaling.}
nGraph-HE2~\cite{boemer2019ngraph2} proposes \textit{lazy rescaling} to reduce the high cost of rescaling operations in CKKS, which, based on their empirical observations, is significantly more expensive than plaintext-ciphertext multiplication. Instead of rescaling after every multiplication, lazy rescaling defers the operation until the end of a full linear layer, thereby amortizing the cost across multiple multiplications. While effective in practice, this method does not explicitly account for the relationship between the latency and ciphertext level, and can thus be regarded as a heuristic optimization. It may be viewed as a trivial form of scale management, which we discuss more systematically in Section~\ref{sssec:scale-mgt-boot-plm}.


\section{Model-Level Optimization}
\label{sec:model}

\begin{figure}[!tb]
    \centering
    \includegraphics[width=\linewidth]{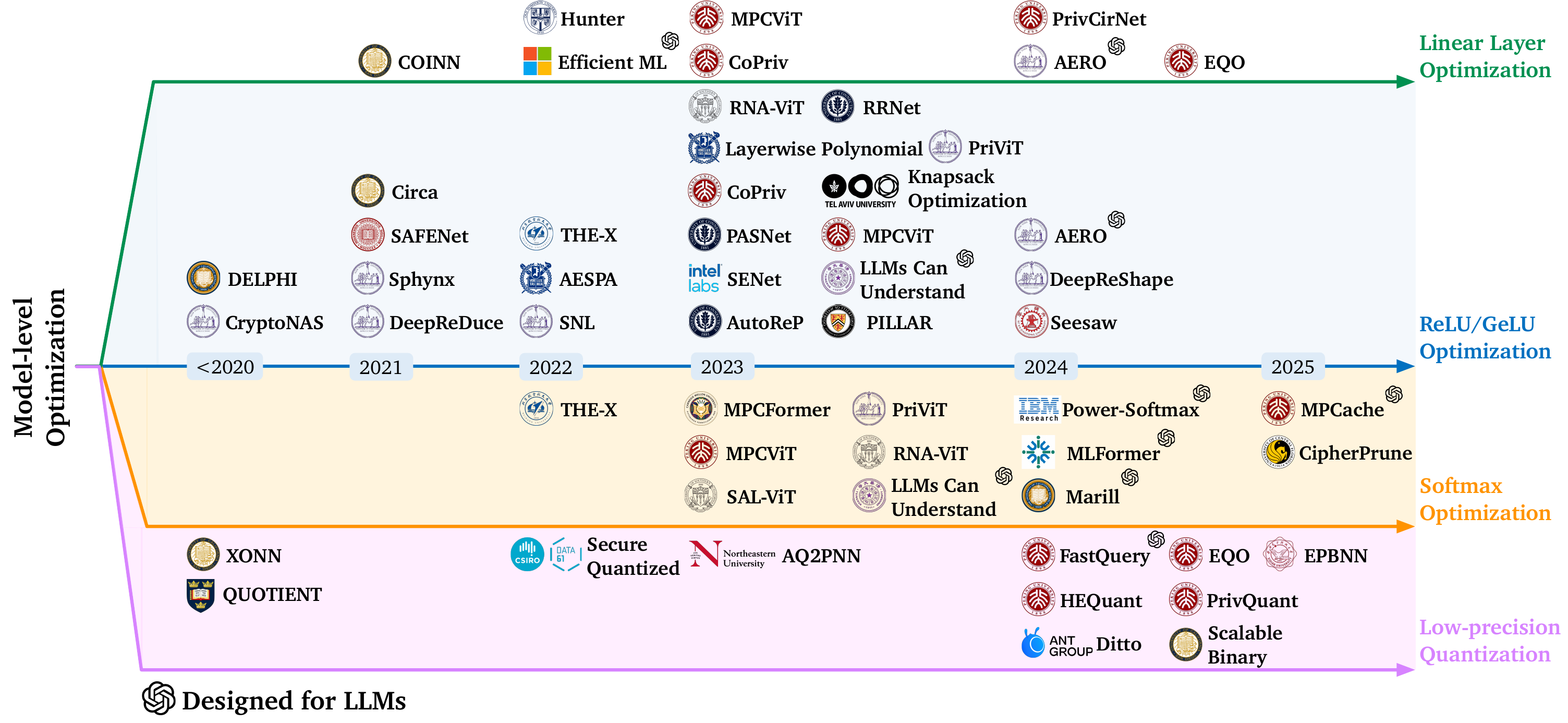}
    \caption{A timeline of existing model-level PPML optimizations for layer and non-linear layers. Model-level optimization heavily relies on model fine-tuning or re-training.}
    \label{fig:timeline_model}
\end{figure}

The model-level optimization mainly focuses on designing a PPML-friendly model architecture, which includes linear layer and non-linear layer optimization.
Linear layer is typically convolution and non-linear layer includes Softmax, ReLU, and GeLU.
We also discuss quantization methods, which benefit both linear and non-linear layers.
Different from protocol-level optimization, model-level optimization typically requires model training or fine-tuning.
We summarize the overall model-level optimizations in Figure \ref{fig:timeline_model}, and mark the methods designed for LLMs that might be very expensive for training or fine-tuning.

\subsection{Linear Layer Optimization}
\label{sec:model_linear}

Linear layers are ubiquitous in various kinds of neural networks.
For CNNs, linear layers mainly involve convolutions, while for Transformers, linear layers mainly involve large matrix multiplications (MatMuls) in attention and feed-forward layers (FFNs).
We summarize the studies in Figure \ref{fig:timeline_model} (green line) and show the details in Table \ref{tab:model_linear}.

\begin{center}
\begin{tcolorbox}[
    colback=myblue!50,
    colframe=myblue,
    width=\linewidth,
    arc=2mm, auto outer arc,
    boxrule=1pt,
    enhanced,
    drop shadow
]
    \raisebox{-0.2cm}{\includegraphics[width=0.55cm]{icon/classify.png}}
    \textbf{Generally, there are two primary ways to improve the efficiency of linear layers:}
    \begin{itemize}
        \item Way 1: Reducing the number of multiplications in each MatMul, e.g., Winograd convolution and circulant convolution.
        \item Way 2: Directly reducing the number of linear layers, e.g., re-parameterization or layer fusion.
    \end{itemize}
\end{tcolorbox}
\end{center}

\begin{table}[!tb]
    \centering
    \caption{Model-level linear layer optimizations.}
    \label{tab:model_linear}
    \resizebox{\linewidth}{!}{
    \begin{tabular}{c|c|c|c|c|>{\columncolor{myblue!70}}c|>{\columncolor{myblue!70}}c|c}
    \toprule
    Method     &  Year & Model & Protocol & Backend & \tabincell{c}{\# Layers \\Reduced} &  \tabincell{c}{\# Mult. Per \\Layer Reduced} & Technique \\
    \hline
    COINN \cite{hussain2021coinn} & 2021 & CNN & SS, OT & EMP-toolkit & \ding{55} & \checkmark & Factorized linear layer \\
    Hunter \cite{hunter}  & 2022 & CNN & HE  & - & \ding{55} & \checkmark &   Structured pruning  \\
    \citet{ganesan2022efficient}  & 2022 & CNN & SS, OT   & EzPC & \ding{55} & \checkmark &  \tabincell{c}{Winograd conv., \\factorized conv.} \\
    CoPriv \cite{zeng2023copriv}  &  2023 & CNN & SS, OT  &  EzPC  & \checkmark & \checkmark & \tabincell{c}{Winograd conv., \\re-parameterization}  \\
    MPCViT \cite{zeng2023mpcvit}  &  2023  & ViT &  SS, OT &  Secretflow & \checkmark & \ding{55} & Layer fusion \\
    PrivCirNet \cite{xu2024privcirnet}  & 2024 & CNN & HE   & SEAL  &  \ding{55} & \checkmark & Block circulant conv. \\
    AERO \cite{jha2024aero}  &  2024 &  GPT-2, Pythia & - & Secretflow  & \checkmark & \ding{55} & Layer fusion  \\
    EQO \cite{zeng2024eqo} & 2024 & CNN  &  SS, OT &  EzPC & \ding{55} & \checkmark &  Quantized Winograd conv. \\
    \bottomrule
    \end{tabular}
    }
\end{table}

For way 1, COINN \cite{hussain2021coinn} proposes factorized MatMul by weight clustering to reduce the number of multiplications.
Hunter \cite{hunter} identifies three HE-friendly structures, i.e., internal structure, external structure, and weight diagonal, and three structured weight pruning modules are correspondingly proposed for dot product and convolution.
\citet{ganesan2022efficient} replaces regular convolution with factorized convolution like depth-wise convolution and channel-wise shuffle to enable information flow across channels. \citet{ganesan2022efficient} also proposes factorized pointwise convolution and uses Winograd convolution for secure inference.
PrivCirNet \cite{xu2024privcirnet} builds a block circulant convolution to simplify convolution computation and further proposes a latency-aware search algorithm for layer-wise block size determination.

For way 2, MPCViT \cite{zeng2023mpcvit} uses NAS to prune the unimportant GeLU in FFN and then fuse two linear layers.
AERO \cite{jha2024aero} also fuses two linear layers in FFN when the ReLU is removed.
More comprehensively, CoPriv \cite{zeng2023copriv} simultaneously reduces the number of linear layers and multiplications in each layer. Specifically, CoPriv involves Winograd convolution and further proposes re-parameterization to fuse adjacent convolution layers in one block after ReLU pruning.
As an optimization of CoPriv, EQO \cite{zeng2024eqo} carefully combines Winograd convolution and quantization to further reduce the cost of convolution.
We believe that optimizing both things simultaneously can bring effective overall efficiency improvement.

\subsection{Non-Linear Layer Optimization}
\label{sec:model_nonlinear}

\subsubsection{ReLU Pruning and Approximation}

ReLU is typically used between two convolution layers in CNNs and defined as
\begin{equation}
    y = \max(0, x).
\end{equation}

However, the max operation is very expensive in PPML.
DeepReDuce \cite{jha2021deepreduce} proposes a three-step method to manually prune redundant ReLUs, which can take a lot of effort and time for optimization.
Circa \cite{ghodsi2021circa} reformulates ReLU as an approximate sign test and proposes a truncation method for the sign test that significantly reduces the cost per ReLU.
AESPA \cite{park2022aespa} uses the low-degree polynomial activation function that exploits the Hermite expansion of ReLU and basis-wise normalization.
\citet{gorski2023securing} finds that ReLU outcomes of nearby pixels are highly correlated, and then proposes to use a single ReLU for each patch. To determine the optimal patch sizes for all layers, this work uses the Knapsack-based optimization strategy to solve this problem.
\citet{lee2023optimizing} proposes a layer-wise approximation method to optimize the polynomial degree for each layer using the dynamic programming algorithm.
PILLAR \cite{diaa2024fast} replaces all ReLUs with a polynomial approximation and proposes polynomial activation regularization to mitigate the accuracy degradation.
DeepReShape \cite{jha2021deepreduce} finds that strategically allocating channels to position the network’s ReLUs based on their criticality to accuracy can simultaneously optimize ReLU and FLOPs efficiency. Therefore, this work proposes ReLU-equalization and ReLU-reuse to enable efficient private inference.

The above works mainly uniformly replace ReLU with other PPML-friendly functions for the whole model.
Another line of research focuses on more fine-grained optimizations such as layer-wise, block-wise, and pixel-wise replacement.
These studies utilize various NAS algorithms to search for the redundant ReLUs and prune these ReLUs at different granularities \cite{mishra2020delphi,ghodsi2020cryptonas,lou2021safenet,cho2021sphynx,cho2022selective,peng2023rrnetrelureducedneuralnetwork,zeng2023copriv,peng2023autorep,kundu2023learning,peng2023pasnet,liseesaw}.
SNL \cite{cho2022selective} proposes a differentiable NAS method to learn the unimportant ReLUs as defined below:
\begin{equation}
    \alpha \cdot \mathrm{ReLU}(x) + (1 - \alpha)\cdot x,
\end{equation}
where $\alpha\in[0, 1]$ is the architecture parameter assigned to each ReLU at pixel-wise or channel-wise granularity.
SENet \cite{kundu2023learning} uses the metric of layer sensitivity to allocate different ReLU budgets for different layers, and then identifies the important ReLUs.
CoPriv \cite{zeng2023copriv} follows the idea of differentiable NAS in SNL and further proposes a communication-aware NAS algorithm to identify the unimportant ReLUs.

\subsubsection{GeLU Pruning and Approximation}
GeLU is widely used in FFNs in Transformers like GPT \cite{brown2020language} and is defined as
\begin{equation}
    \mathrm{GeLU}(x) = \frac{x}{2}(1+\mathrm{erf}(\frac{x}{\sqrt 2})),
\end{equation}
where $\mathrm{erf}(\cdot)$ denotes the Gaussian error function. GeLU is usually approximated with tanh function as
\begin{equation}
    \mathrm{GeLU}(x) = \frac{x}{2}(1 + \mathrm{tanh}(\sqrt{\frac{2}{\pi}}(x + 0.044715x^3))).
\end{equation}

As observed, GeLU is inefficient for PPML due to the non-linear error function, i.e., tanh and the square root.

\begin{center}
\begin{tcolorbox}[
    colback=myblue!50,
    colframe=myblue,
    width=\linewidth,
    arc=2mm, auto outer arc,
    boxrule=1pt,
    enhanced,
    drop shadow
]
    \raisebox{-0.2cm}{\includegraphics[width=0.55cm]{icon/classify.png}}
    \textbf{Generally, there are two primary ways to reduce the overhead of GeLU:}
    \begin{itemize}
        \item Way 1: Replacing GeLU with efficient alternatives such as ReLU and quadratic function through model training.
        \item Way 2: Approximating GeLU with high-order polynomials without training.
    \end{itemize}
\end{tcolorbox}
\end{center}

Some works simply replace GeLU with ReLU to reduce the overhead \cite{chen2022x,liu2023llms}, and RNA-ViT \cite{chen2023rna} replaces GeLU with LeakyReLU to improve the model performance.
AERO \cite{jha2024aero} even more directly prunes the deeper layers due to the observation of their redundancy.
MPCFormer \cite{li2022mpcformer} and Ditto \cite{wu2024ditto} approximate GeLU with a quadratic polynomial as 
\begin{equation}
    \mathrm{GeLU}(x)=0.125x^2+0.25x+0.5.
\end{equation}

MPCViT \cite{zeng2023mpcvit} and PriViT \cite{dhyani2023privit} propose to use differentiable NAS to prune unimportant GeLU. MPCViT further proposes to fuse two adjacent linear layers after pruning to further improve the efficiency.
The NAS algorithm is formulated as
\begin{equation}
    \alpha \cdot \mathrm{GeLU}(x) + (1 - \alpha)\cdot x,
\end{equation}
where $\alpha\in[0, 1]$ is the architecture parameter assigned to each GeLU, indicating the importance of each GeLU.

\textbf{High-order polynomial approximation without training.}
Different from the above approximations which require training, high-order polynomial approximations can be directly applied to inference; however, introduce a larger overhead to compute the high-order polynomials.
PUMA \cite{dong2023puma} observes that GeLU is almost linear on the two sides, i.e., $\mathrm{GeLU}(x) \approx 0$ for $x < -4$ and $\mathrm{GeLU}(x) \approx x$ for $x > 3$, and proposes a piece-wise polynomial approximation with four segments as follows:
\begin{equation}
    \mathrm{GeLU}(x) = 
    \left\{
    \begin{array}{ll} 
    0, & \text{if } x < -4 \\
    F_0(x), & \text{if } -4 \le x < -1.95 \\
    F_1(x), & \text{if } -1.95 \le x \le 3 \\ 
    x, & \text{if } x > 3 \\
    \end{array} 
    \right.,
\end{equation}
where $F_0(x)$ and $F_1(x)$ are 3-order and 6-order polynomials, respectively computed by the \texttt{numpy.polyfit} tool.
BumbleBee \cite{lu2023bumblebee} proposes a similar approximation method as follows:
\begin{equation}
    \mathrm{GeLU}(x) = 
    \left\{
    \begin{array}{ll} 
    -\epsilon, & \text{if } x < -5 \\
    F_0(x), & \text{if } -5 \le x < -1.97 \\
    F_1(x), & \text{if } -1.97 \le x \le 3 \\ 
    x-\epsilon, & \text{if } x > 3 \\
    \end{array} 
    \right.,
\end{equation}
where $\epsilon$ denotes a small value such as $1e-5$, $F_0(x)$ and $F_1(x)$ are 3-order and 6-order polynomials, respectively computed by the \texttt{numpy.polyfit} tool.
Besides the linearity of GeLU on the two sides, BOLT \cite{pang2024bolt} observes the symmetry of $x\cdot \mathrm{erf}(x/\sqrt 2)$ and proposes to approximate GeLU for half of the input range, i.e., positive input values as follows:
\begin{equation}
    \mathrm{GeLU}(x) = 
    \left\{
    \begin{array}{ll} 
    x, & \text{if } x > 2.7 \\
    a|x|^4+b|x|^3+c|x|^2+d|x|+e+0.5x , & \text{if } |x| \le 2.7 \\
    0, & \text{if } x < -2.7
    \end{array} 
    \right.,
\end{equation}
where the coefficients $a, b, c, d, e$ are computed with the Remez method.

We summarize the studies of ReLU and GeLU optimizations in Table \ref{tab:model_relu_gelu}.
\begin{table}[!tb]
    \centering
    \caption{Model-level non-linear ReLU and GeLU optimizations.}
    \label{tab:model_relu_gelu}
    \resizebox{\linewidth}{!}{
    \begin{tabular}{c|c|c|c|c|>{\columncolor{myblue!70}}c|>{\columncolor{myblue!70}}c|c}
    \toprule
    Method     &  Year & Model & Protocol & Backend & NAS & Granularity & Technique \\
    \hline
    Delphi \cite{mishra2020delphi}  & 2020 & CNN & GC & fancy-garbling & \checkmark & Layer & Population algorithm \\
    CryptoNAS \cite{ghodsi2020cryptonas} & 2020 & CNN & GC & MiniONN & \checkmark &  Layer &  Macro-search and ReLU shuffling \\
    DeepReDuce \cite{jha2021deepreduce} & 2021 & CNN & GC & Delphi & \ding{55} & 
 Layer & Manually pruning  \\
    SAFENet \cite{lou2021safenet} & 2021 & CNN & GC & fancy-garbling & \checkmark & Channel & Multiple-degree Poly. approximation \\
    Sphynx \cite{cho2021sphynx} & 2021 & CNN & GC & Delphi & \checkmark & Block & Micro-search algorithm \\
    Circa \cite{ghodsi2021circa} & 2021 & CNN & GC & Delphi & \ding{55} & Model  & Piece-wise linear and sign function \\
    SNL \cite{cho2022selective}  & 2022  & CNN & GC & Delphi & \checkmark & Pixel, Channel &  Differentiable search \\
    AESPA \cite{park2022aespa} & 2022 & CNN & GC & Delphi & \ding{55} & Model &  Hermite ReLU expansion \\
    THE-X \cite{chen2022x} & 2022 & Bert  & - & - & \ding{55} & Model & Replace GeLU with ReLU \\
    RRNet \cite{peng2023rrnetrelureducedneuralnetwork} & 2023 & CNN &  SS, OT & - & \checkmark & Layer & Hardware-aware search \\
    MPCFormer \cite{li2022mpcformer}  & 2023 & Bert, Roberta & SS, OT & CrypTen  & \ding{55} & Model &  Replace GeLU with quadratic poly. \\
    MPCViT \cite{zeng2023mpcvit}  & 2023  & ViT & SS, OT  & Secretflow & \checkmark & Layer & Latency-aware search \\
    PriViT \cite{dhyani2023privit} &  2023 & ViT & SS, OT & Secretflow & \checkmark & Layer & Differentiable search  \\
    CoPriv \cite{zeng2023copriv}  & 2023 & CNN & SS, OT  & EzPC  & \checkmark & Block & Communication-aware search  \\
    AutoRep \cite{peng2023autorep} & 2023 & CNN & SS, OT & CrypTen & \checkmark & Pixel & Distribution-aware poly. approx. \\
    SENet \cite{kundu2023learning} & 2023 & CNN & GC & Delphi & \checkmark & Layer, pixel & Sensitivity-based pruning \\
    PASNet \cite{peng2023pasnet} & 2023 & CNN & \cite{garay2007practical} & \cite{garay2007practical} & \checkmark & Layer & Hardware-aware search   \\
    \citet{gorski2023securing} & 2023 & CNN & 3PC compare & SecureNN & \ding{55} & Patch, channel & Knapsack-based pruning \\
    \citet{liu2023llms} & 2023 & Bert, Roberta & SS, OT  & EzPC & \ding{55} & Model & Replace GeLU with ReLU \\
    \citet{lee2023optimizing} & 2023 & CNN & HE (RNS-CKKS) & Lattigo & \ding{55} & Layer  & \tabincell{c}{Layer-wise degree poly. \\with dynamic programming} \\
    PILLAR \cite{diaa2024fast} & 2023  &  CNN  & SS, OT & CrypTen & \ding{55} & Model  & Polynomial regularization  \\
    RNA-ViT \cite{chen2023rna} & 2023 & ViT & SS, OT & CrypTen & \ding{55} & Model & Replace GeLU with LeakyReLU \\
    AERO \cite{jha2024aero} & 2024 & GPT-2, Pythia  & - & Secretflow & \ding{55} 
 & Layer & Deeper-layer pruning  \\
    Seesaw \cite{liseesaw} &  2024 & CNN & GC & Delphi & \checkmark & Block & Bi-level search \\
    DeepReShape \cite{jha2024deepreshaperedesigningneuralnetworks} & 2024 & CNN & GC & Delphi & \ding{55} & Stage & ReLU equalization and reuse \\
    Ditto \cite{wu2024ditto} & 2024 & Bert, GPT-2 & RSS  & Secretflow & \ding{55} & Model & Replace GeLU with quadratic poly.  \\
    \bottomrule
    \end{tabular}
    }
\end{table}

\subsubsection{Softmax Approximation}
\label{sssec:softmax-app}
For Transformers, the non-linear function Softmax is a key component for attention normalization, which is formulated as
\begin{equation}
    \mathrm{Softmax}(x_i) = \frac{e^{x_i-x_{\max}}}{\sum^N_{j=1}e^{x_i-{x_{\max}}}},
\end{equation}
where $N$ denotes the number of tokens and $x_{\max}$ denotes the maximum value of the attention weight for the sake of numerical stability \cite{wang2022characterization}.

As can be observed, computing Softmax involves different PPML-inefficient operators, including exponential, max, and division.
In recent years, studies have mainly focused on getting rid of these unfriendly operators as much as possible while maintaining the model performance.

\begin{center}
\begin{tcolorbox}[
    colback=myblue!50,
    colframe=myblue,
    width=\linewidth,
    arc=2mm, auto outer arc,
    boxrule=1pt,
    enhanced,
    drop shadow
]
    \raisebox{-0.2cm}{\includegraphics[width=0.55cm]{icon/classify.png}}
    \textbf{Generally, there are three primary ways to reduce the overhead of Softmax:}
    \begin{itemize}
        \item Way 1: Replacing PPML-inefficient operators, including exponential, max, and division with efficient operators through model training.
        \item Way 2: Directly reducing the number of Softmax, e.g., head pruning/merging and KV cache compression.
        \item Way 3: Replacing PPML-inefficient exponential with high-order polynomials without training.
    \end{itemize}
\end{tcolorbox}
\end{center}

Most of the works focus on way 1 by removing the inefficient operators.
MPCFormer \cite{li2022mpcformer} and SecFormer \cite{luo2024secformer} uniformly replaces exponential with quadratic function as $\mathrm{Softmax}(x)\approx(x+c)^2/\sum(x+c)^2$ where $c$ is a constant. 
MPCViT \cite{zeng2023mpcvit} proposes to replace exponential with ReLU and further use scaling attention for ``unimportant'' heads using latency-aware neural architecture search (NAS).
Following MPCViT, PriViT \cite{dhyani2023privit} directly replaces the whole Softmax with a quadratic function as $\mathrm{Softmax}(x)\approx x^2/N$ where $N$ is the token number, and combines it with original Softmax using NAS.
SAL-ViT \cite{zhang2023sal} further proposes to combine self-attention and external attention \cite{guo2022beyond} using NAS to improve MPCViT.
RNA-ViT \cite{chen2023rna} first uses linear layers to reduce the dimension and then uses a high-order polynomial (HOP) to approximate Softmax as $\mathrm{Softmax}(x)=(x+c)^4/(\sum(x+c)^4+\epsilon)$ where $c$ is a constant and $\epsilon$ is a small positive value to avoid a zero denominator.
MLFormer \cite{liu2025mlformer} considers this issue from the perspective of linear attention and proposes scaled linear attention to avoid Softmax.
Power-Softmax \cite{zimerman2024power} proposes $\mathrm{Softmax}(x)\approx x^p/\sum x^p$ where $p$ is a positive even number.
To circumvent the use of vector-wise Softmax, \citet{zimerman2023converting} employs an element-wise function like a quadratic function or a high-degree polynomial.

For way 2, Marill \cite{rathee2024mpc} proposes using head pruning and merging in PPML to directly reduce the number of Softmax, thus reducing the inference overhead. Note that the process of deciding which heads to prune or merge is done offline \cite{bian2021attention}, so there is no additional inference overhead. MPCache \cite{zeng2025mpcache} for the first time incorporates the KV cache eviction algorithm with PPML-friendly optimizations to drastically reduce the KV cache, thereby lowering the Softmax overhead.
CihperPrune \cite{hou2023ciphergpt} proposes encrypted token pruning to remove unimportant tokens layer by layer and assigns lower-degree polynomials to less important tokens after pruning.

\textbf{High-order polynomial approximation without training.}
PUMA \cite{dong2023puma}, CipherGPT \cite{hou2023ciphergpt}, and BumbleBee \cite{lu2023bumblebee} approximate exponential using Taylor series as follows:
\begin{equation}
    \mathrm{exp}(x) = 
    \left\{
    \begin{array}{ll} 
    0, & \text{if } x \le m \\
    (1+\frac{x}{2^t})^{2^t}, & \text{if } m \le x \le 0 \\
    \end{array} 
    \right.,
\end{equation}
where $m$ and $t$ are set to different values in these works with different efficiency and performance trade-offs.
Specifically, BumbleBee sets $a=-13, t=6$, CipherGPT sets $a=-16$, PUMA sets $a=-14, t=5$.

\textbf{LUT-based evaluation.}
Some studies directly utilize LUTs to evaluate the non-linear Softmax, which pre-compute and store the input-output pairs in the table to avoid costly real-time computation.
Iron \cite{hao2022iron} follows CryptFlow2 \cite{rathee2020cryptflow2} and SiRNN \cite{rathee2021sirnn} and uses OT-based LUT to evaluate the exponential and reciprocal.
Pika \cite{wagh2022pika} uses large LUTs for exponential and reciprocal, but large LUTs are very inefficient.
SIGMA \cite{gupta2023sigma} further proposes to minimize the size of LUTs with lower bit widths while maintaining accuracy.

\begin{table}[!tb]
    \centering
    \caption{Model-level non-linear Softmax optimizations.}
    \label{tab:model_softmax}
    \resizebox{\linewidth}{!}{
    \begin{tabular}{c|c|c|c|c|>{\columncolor{myblue!70}}c|>{\columncolor{myblue!70}}c|c}
    \toprule
    Method     &  Year & Model & Protocol & Backend & NAS & Granularity & Technique \\
    \hline
    THE-X \cite{chen2022x} & 2022 & Bert & HE (CKKS) & SEAL & \ding{55} & Model & Extra estimation model \\
    MPCFormer \cite{li2022mpcformer}  & 2023 & Bert, Roberta & SS, OT & CrypTen  & \ding{55} & Model & Replace $e^x$ with $x^2$  \\
    MPCViT \cite{hussain2021coinn} & 2023 & ViT & SS, OT & Secretflow  & \checkmark & Head & \tabincell{c}{Replace $e^x$ to ReLU \\and scaling attention} \\
    SAL-ViT \cite{zhang2023sal} & 2023  & ViT & SS, OT &  CrypTen & \checkmark & Head  & \tabincell{c}{Replace self attention \\with external attention} \\
    PriViT \cite{dhyani2023privit} & 2023 & ViT & SS, OT & Secretflow & \checkmark  & Row  & Replace Softmax with Square \\
    RNA-ViT \cite{chen2023rna} & 2023 & ViT & SS, OT & CrypTen & \ding{55} & Model &  Higher-order poly. approx. \\
    MLFormer \cite{liu2025mlformer} & 2024 & ViT & SS, OT & CrypTen & \ding{55} & Model & Scaled linear attention  \\
    Marill \cite{rathee2024mpc} & 2024 & \tabincell{c}{LLaMA, \\Deepseek Coder} & GC &  SecureML & \ding{55} & Head &  Head pruning and merging \\
    Power-Softmax \cite{zimerman2024power} & 2024 & Pythia, Roberta & HE (CKKS) & HEaaN & \ding{55} & Model & Poly. approx. and Lipschitz division \\
    \citet{zimerman2023converting} & 2024 & ViT, Bert & HE (CKKS) & HEaaN & \ding{55} &  Model & HE-friendly poly. approx. \\
    SecFormer \cite{luo2024secformer} & 2024 & Bert & SS, OT & CrypTen  & \ding{55} & Model & Replace $e^x$ with $x^2$  \\
    MPCache \cite{zeng2025mpcache} & 2025 & GPT, LLaMA & SS, OT & Secretflow & \ding{55} & Head & KV cache compression \\
    CipherPrune \cite{zhang2025cipherprune} & 2025  & GPT, Bert & SS, OT, HE & EzPC &  \checkmark & Layer & Token pruning and poly. approx. \\
    \bottomrule
    \end{tabular}
    }
\end{table}

\begin{center}
\begin{tcolorbox}[
    colback=orange!8,
    colframe=orange!20,
    width=\linewidth,
    arc=2mm, auto outer arc,
    boxrule=1pt,
    enhanced,
    drop shadow
]
    \raisebox{-0.2cm}{\includegraphics[width=0.5cm]{icon/light.png}}
    \textbf{Takeaways of PPML-friendly non-linear layers:}
    \begin{itemize}
        \item As observed in these studies, approximation or pruning at model-wise granularity can hurt the model performance significantly. To alleviate this problem, more and more studies focus on fine-grained approximations like layer-wise and head-wise using NAS algorithms.
        \item Replacing non-linear layers with simpler operators or low-order polynomials requires training to recover the model performance. In contrast, when approximating non-linear layers using higher-order polynomials ($\ge 3$-order) does not necessarily require training, but suffers from higher overhead for evaluating the polynomial.
    \end{itemize}
\end{tcolorbox}
\end{center}

\subsection{Low-Precision Quantization}
\label{sec:quant}

Quantization is a widely used technique that compresses the activations or weights to low bit widths for efficient training or inference.
Quantization benefits both linear and non-linear layers.
However, directly applying quantization to PPML is a non-trivial practice for both SS-based and HE-based frameworks.
We summarize the quantization methods designed for PPML in Table \ref{tab:model_quant}.

\begin{table}[!tb]
    \centering
    \caption{Model-level PPML-friendly quantization optimizations.}
    \label{tab:model_quant}
    \resizebox{\linewidth}{!}{
    \begin{tabular}{c|c|c|c|c|>{\columncolor{myblue!70}}c|c}
    \toprule
    Method     &  Year & Model & Protocol & Backend  & Mixed-Precision & Technique \\
    \hline
    XONN \cite{riazi2019xonn}  & 2019  & CNN & GC & Customization & \ding{55} & Binary quant. \\
    COINN \cite{hussain2021coinn} & 2021 & CNN & SS, OT, GC &  EMP-toolkit, TinyGarble2 & \checkmark & Genetic algorithm  \\
    PrivQuant \cite{xu2024privquant} & 2023 & CNN & SS, OT & EzPC & \checkmark & Communication-aware quant. \\
    EQO \cite{zeng2024eqo} & 2024 &  CNN & SS, OT & EzPC & \checkmark & \tabincell{c}{Communication-aware Winograd quant.,\\ bit re-weighting for outliers} \\
    Ditto \cite{wu2024ditto} & 2024 & Bert, GPT-2 & RSS & Secretflow & \ding{55} & Static dyadic quant. \\
    FastQuery \cite{lin2024fastquery}  &  2024  & LLaMA-2 & HE & OpenCheetah & \checkmark &  Communication-aware embedding table quant. \\
    \bottomrule
    \end{tabular}
    }
\end{table}

\subsubsection{SS-Friendly Algorithm}
XONN \cite{riazi2019xonn} is a very early work that attempts to reduce the cost of the GC protocol.
XONN uses binary neural networks (BNNs) which quantize the activations and weights to binary values to replace the costly multiplications with XNOR operations.
COINN \cite{hussain2021coinn} enforces the scale to be powers of 2, which allows us to compute the costly scale operation using logical shifts in GC with nearly zero cost.
PrivQuant \cite{xu2024privquant} proposes to use mixed-precision quantization based on communication-aware Hessian information \cite{dong2019hawq} for a better trade-off between model accuracy and efficiency. 
PrivQuant finds that direct quantization even increases the inference overhead due to expensive bit width conversions, including extension and re-quantization. PrivQuant then proposes a series of graph-level optimizations like protocol fusion and most significant bit (MSB)-known optimization to reduce the overhead.
EQO \cite{zeng2024eqo} further combines efficient Winograd convolution and mix-precision quantization, but observes that Winograd transformations can introduce weight outliers and extra costly bit width extensions. To solve these problems, EQO proposes a bit re-weighting strategy to accommodate the outliers and graph-level protocol fusion to remove the extra extensions.
Ditto \cite{wu2024ditto} adopts static dyadic quantization to avoid dynamically computing scale during inference and proposes type conversion protocols for efficient bit width conversion.

\subsubsection{HE-Friendly Algorithm}

FastQuery \cite{lin2024fastquery} features a communication-aware embedding table quantization algorithm that assigns different bit widths for different channels.
FastQuery leverages the one-hot nature of queries and the low-precision embedding table to reduce the accumulation bit width. 
To further reduce the communication for sub-13-bit quantization, FastQuery proposes a novel element packing algorithm to squeeze the embedding table dimension.

\begin{center}
\begin{tcolorbox}[
    colback=orange!8,
    colframe=orange!20,
    width=\linewidth,
    arc=2mm, auto outer arc,
    boxrule=1pt,
    enhanced,
    drop shadow
]
    \raisebox{-0.2cm}{\includegraphics[width=0.5cm]{icon/light.png}}
    \textbf{Takeaways of PPML-friendly quantization:}
    \begin{itemize}
        \item Directly combining quantization and PPML is a non-trivial practice. Instead, the combination usually requires model/algorithm and PPML protocol co-optimization to enable end-to-end efficiency improvement.
        \item Although there are several SS-friendly quantization algorithms, the algorithm designed for HE is still scarce. We believe the main reason is that quantization is not simple for HE, requiring very refined quantization strategies, modulus selection, encoding methods, etc.
    \end{itemize}
\end{tcolorbox}
\end{center}
\section{System-Level Optimization}
\label{sec:system}
System-level optimization is essential for accelerating PPML, as the computational bottleneck typically resides in the HE component, whether used alone or in combination with MPC. Protocol-level encoding optimizations reveal challenges of designing efficient HE-style workflows. While such techniques can reduce the number of homomorphic operations, they offer limited relief from the intrinsic computational overhead of each operation. Compiler technologies help bridge the semantic gap between high-level ML workloads and low-level HE primitives, while GPUs, as the most accessible accelerators, help narrow the performance gap between encrypted and plaintext computation. This section highlights recent advances in HE compilers and GPU optimizations, which together enable scalable and efficient PPML systems.

\subsection{Compiler}
\label{sec:compiler}

\begin{figure}[!tb]
    \centering
    \includegraphics[width=\linewidth]{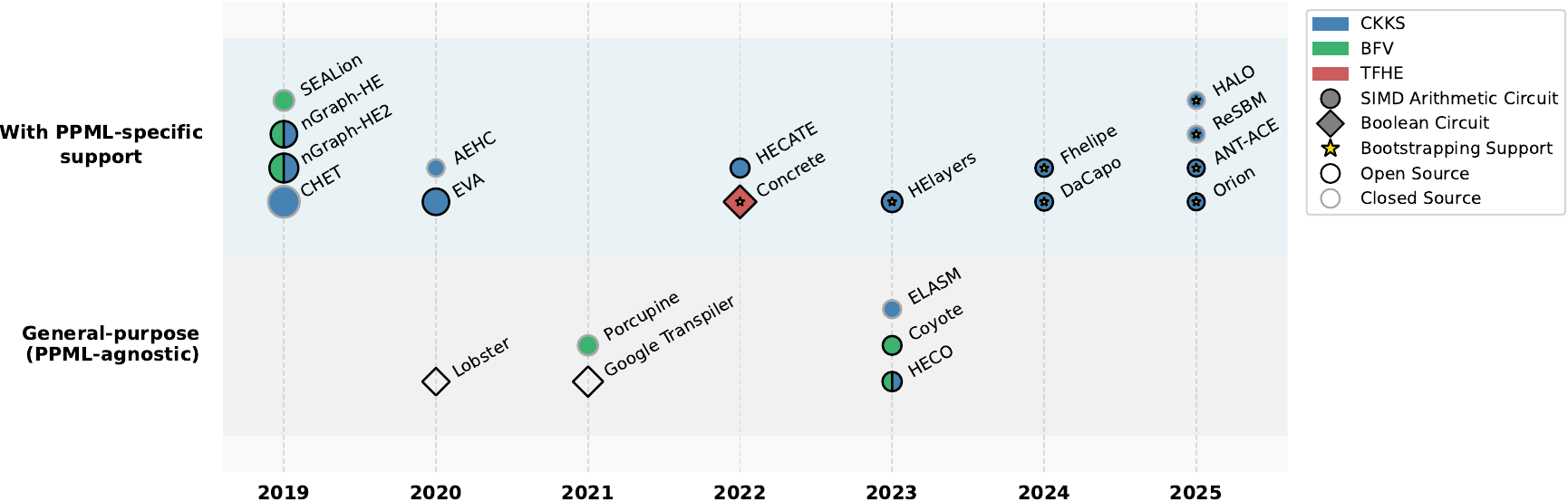}
    \caption{Timeline of HE Compilers. PPML-specific compilers are positioned in the upper region (Classification is based on whether the corresponding paper introduces an ML-friendly IR or DSL, or includes evaluations on PPML tasks).}
    \label{fig:compiler-roadmap}
\end{figure}

We categorize existing HE compilers into two groups based on their target workloads: \textbf{general-purpose} compilers and those \textbf{with PPML-specific support}, as illustrated in Figure~\ref{fig:compiler-roadmap}. We first motivate the need for HE compilers, especially for PPML, and then review core techniques.

\subsubsection{The Need for HE Compiler}
Writing efficient FHE programs manually is extremely difficult, due to the mathematical structure and the practical demands of real-world workloads.

At a general level, programming in HE involves several unique challenges:
\begin{enumerate}
    \item \textbf{Dataflow constraints}: HE programs must be expressed as static circuits or dataflow graphs, without native control flow constructs like branching or loops.
    \item \textbf{Limited instruction set}: FHE schemes support a small set of homomorphic operations (e.g., addition, multiplication, rotation). Expressing complex functions often requires mathematical approximations.
    \item \textbf{Performance tuning}: Achieving efficient execution requires careful management of noise growth, message precision and encryption parameters, which is non-trivial even for experts.
\end{enumerate}

\begin{center}
\begin{tcolorbox}[
    colback=myblue!50,
    colframe=myblue,
    width=\linewidth,
    arc=2mm, auto outer arc,
    boxrule=1pt,
    enhanced,
    drop shadow
]
    \raisebox{-0.2cm}{\includegraphics[width=0.55cm]{icon/classify.png}}
    \textbf{When it comes to \textit{PPML workloads}, additional difficulties emerge:}
    \begin{itemize}
        \item Neural networks involve many \textbf{non-linear activations}, making them hard to express efficiently in HE.
        \item Optimizing \textbf{data layout} for SIMD encoding is crucial for large tensor operations in the deeply nested computation graphs of neural networks, but doing so manually is extremely challenging.
        \item For CKKS---the most widely used FHE scheme for ML---proper \textbf{scale management} and \textbf{bootstrapping placement} are essential for both correctness and efficiency, and remain non-trivial even for experts.
    \end{itemize}
\end{tcolorbox}
\end{center}

\subsubsection{Packing Optimization}
Efficient packing is critical for leveraging the SIMD capabilities provided by RLWE-based schemes (e.g., BFV, BGV, CKKS). Several operator-level, expert-designed encoding strategies have been discussed in Section~\ref{sssec:he-based-protocol}. Compiler-level packing primarily focuses on \textit{SIMD encoding} (or vectorization), which offers more flexible data movement semantics (i.e., \textsf{Rotate} operations). Over time, these compiler techniques have evolved from domain-specific heuristics to more general, automated strategies that jointly optimize vectorization and data layout.

\textbf{Expert-driven and fixed-layout packing.}
Early systems such as CHET~\cite{dathathri2019chet} automated the selection of encryption parameters and data layouts for tensor circuits using heuristics tailored to PPML workloads. CHET introduced the \textit{CipherTensor} abstraction, which encapsulates a fixed set of supported layouts, each associated with a dedicated set of algorithmic kernels. EVA~\cite{dathathri2020eva} extended this approach by introducing a domain-specific language (DSL) called Encrypted Vector Arithmetic (EVA). Nevertheless, both CHET and EVA still required developers to manually restructure applications into EVA’s batched execution model, limiting ease-of-use and efficiency.

\textbf{Flexible layouts and automatic vectorization.}
To reduce reliance on manual layout design, later systems pursued more flexible and automated strategies. AHEC~\cite{chen2020ahec} extended the \textit{Tile DSL} from PlaidML~\cite{plaidml} to represent computations as element-wise operations suitable for vectorization, but delegated layout decisions entirely to the DSL, missing FHE-specific optimizations.
Porcupine~\cite{cowan2021porcupine} adopted sketch-based synthesis to 
generate rotation-efficient 
kernels, though it requires user-provided sketches. Coyote~\cite{malik2023coyote} removed this barrier, enabling fully automatic vectorization for general programs by optimizing layout and rotations using a cost model to minimize latency and ciphertext noise growth. HECO~\cite{viand2023heco} 
proposed a batching abstraction called \textit{BatchedSecrets}, which avoids masking and supports operation chaining despite introducing some rotate-and-sum overhead.

Fixed row- or column-major layouts often lead to sub-optimal rotation patterns. To address this, recent compilers such as HELayer~\cite{aharoni2020helayers} and Fhelipe~\cite{krastev2024tensor} introduce expressive layout abstractions that open up broader optimization opportunities for automatic packing. HELayer proposes a \textit{tile tensor} design, which generalizes layout support using a unified set of algorithms. Fhelipe introduces a flexible tensor layout framework that supports arbitrary dimension orders, interleaving, and gaps. It explores a wider optimization space through an analytical layout selection algorithm that minimizes layout conversions and maximizes packing density. 
FlexHE~\cite{yu2024flexhe} further enables generating and optimizing coefficient encoding-based 2PC-compatible HE kernels from einsum-like DNN representation. 

\subsubsection{Scale Management and Bootstrapping Placement}
\label{sssec:scale-mgt-boot-plm}
In CKKS~\cite{cheon2017homomorphic}, real-valued vectors are encoded into plaintext polynomials, where each value is scaled to an integer (the significand) with a fixed-point \textit{scale}. RNS-CKKS~\cite{cheon2018full} encrypts these polynomials under a ciphertext modulus composed of multiple RNS limbs, and the total \textit{scale capacity} equals their product.
Homomorphic multiplication inflates the scale, necessitating \textsf{Rescale} to keep it within capacity. In addition, ciphertexts must be at the same \textit{level} (i.e., same number of limbs), requiring \textsf{ModSwitch} when levels mismatch. As computation proceeds, accumulated scale can exceed capacity, triggering the need for \textit{bootstrapping}~\cite{cheon2018bootstrapping}.

\begin{figure}[!tb]
    \centering
    \includegraphics[width=\linewidth]{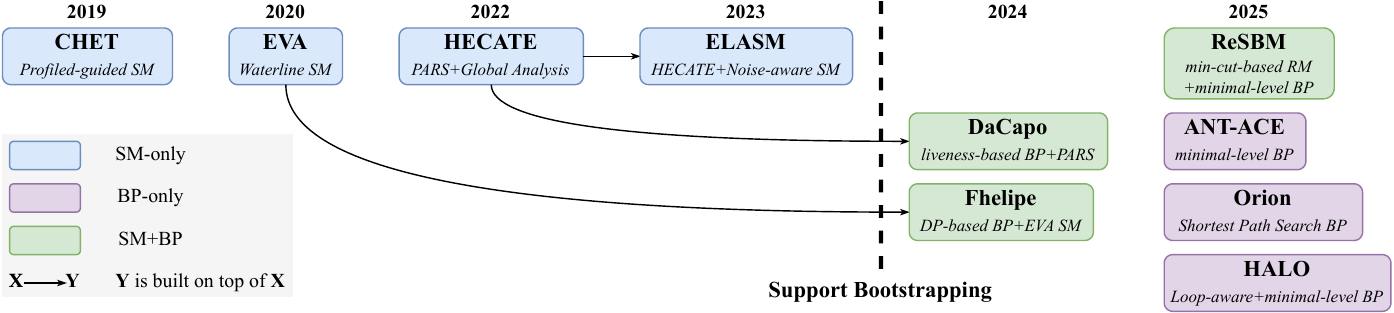}
    \caption{Roadmap of CKKS compilers focusing on scale management (SM) and bootstrapping placement (BP) optimizations. For each node, the top half shows the \textbf{compiler name} in bold, and the bottom half summarizes its \textit{key techniques} in italics.}
    \label{fig:smbp-compiler-roadmap}
\end{figure}

In general, effective \textit{scale management} involves (1) selecting appropriate scaling factors and (2) determining the optimal insertion points for \textsf{Rescale} and \textsf{ModSwitch} operations across the computational graph to minimize overall latency. Optimal \textit{bootstrapping placement} builds upon this scale management strategy, with the shared goal of latency reduction, particularly for DNN workloads. The development roadmap of PPML-oriented CKKS compilers that address these challenges is summarized in Figure~\ref{fig:smbp-compiler-roadmap}.

\textbf{Scale management-only compilers.}
CHET~\cite{dathathri2019chet} provides only a profile-guided optimization for selecting scaling factors, without addressing the placement of \textsf{Rescale} and \textsf{ModSwitch} operations. EVA~\cite{dathathri2020eva} introduces the first coarse-grained scale management strategy by defining a \textit{waterline}, which is the minimum scale required to preserve message integrity. It performs \textsf{Rescale} operations reactively whenever the rescaled value remains above the waterline, passively reducing the ciphertext scale as much as possible. The insight behind EVA is that smaller ciphertext scales enable the use of smaller HE parameters, which in turn improves overall performance. HECATE~\cite{lee2022hecate} further advances scale management by recognizing the cascading impact of \textsf{Rescale} on the performance of subsequent operations. Since the computational complexity of homomorphic operations is directly affected by the ciphertext level, HECATE introduces a fine-grained, performance-aware scale management. It proactively inserts \textsf{Rescale} (referred to as \textsf{downscale} or PARS) to expand the optimization space for minimizing end-to-end latency. 
Unlike EVA’s two-phase approach, which first performs rescaling and then aligns levels, HECATE considers \textsf{Rescale} and \textsf{ModSwitch} operations holistically through hill-climbing-based space exploration. Building on HECATE, ELASM~\cite{lee2023elasm} introduces the first noise-aware scale management strategy. It balances latency and noise growth by configuring a more robust waterline to prevent scale underflow, which could otherwise affect the quality of service in PPML workloads.

\textbf{Bootstrapping-capable compilers.}
The multiplicative depth required by DNNs often exceeds the limits of leveled HE, necessitating the most expensive FHE operation---bootstrapping---to enable more computation. Recent HE compilers have begun to support automatic bootstrapping placement to address this challenge. Contrary to bootstrapping increases the latency of subsequent operations. Therefore, effective bootstrapping management must be tightly integrated with scale management, especially given the NP-hard nature of optimal bootstrapping placement~\cite{paindavoine2015minimizing}.

HECATE and ELASM determine optimal scale management plans by evaluating the cumulative cost of candidate plans using hill-climbing-based space exploration. However, this global optimization incurs high compilation overhead; for instance, compiling LeNet-5 takes over 300 seconds~\cite{lee2022hecate,lee2023elasm}, rendering the approach impractical for larger models. As a result, recent bootstrapping-capable compilers typically avoid global analysis for scale management. 

Fhelipe~\cite{krastev2024tensor} adopts a dynamic programming-based approach to generate bootstrapping plans based on EVA’s waterline-driven scale management. DaCapo~\cite{cheon2024dacapo} leverages liveness analysis to guide bootstrapping placement based on PARS~\cite{lee2022hecate} for scale management. However, both Fhelipe and DaCapo elevate ciphertexts to the maximum level without considering the latency implications of bootstrapping on subsequent operations, often resulting in sub-optimal plans. ReSBM~\cite{liu2025resbm} addresses this limitation by elevating ciphertexts only to the minimal required level, guided by its proposed min-cut-based scale management strategy. It is a region-based compiler that tightly coordinates scale management and bootstrapping placement for both compile-time and runtime efficiency. HALO~\cite{cheon2025halo} and ANT-ACE~\cite{li2025ant} also adopt this minimal-level bootstrapping strategy but do not explicitly elaborate on their scale management approaches. Notably, HALO is the first to propose loop-aware bootstrapping placement, enabling optimized code generation within non-unrolled loops. 
Orion~\cite{ebel2023orion} allows more flexible bootstrapping initiation points within the circuit. It formulates the placement problem as a shortest-path search over a DAG and incorporates network-specific heuristics to guide the search. However, Orion does not integrate scale management with bootstrapping; instead, it relies on a conservative, manually configured scale management~\cite{bossuat2021efficient,kim2022approximate}.

\begin{center}
\begin{tcolorbox}[
    colback=orange!8,
    colframe=orange!20,
    width=\linewidth,
    arc=2mm, auto outer arc,
    boxrule=1pt,
    enhanced,
    drop shadow
]
    \raisebox{-0.2cm}{\includegraphics[width=0.5cm]{icon/light.png}}
    \textbf{Takeaways of PPML-oriented HE compilers:}
    \begin{itemize}
        \item Manually developing performant FHE programs is error-prone and labor-intensive due to limited operations and complex trade-offs among correctness, latency, and noise. These challenges are further exacerbated in PPML workloads, motivating PPML-specific HE compilers.
        \item Compiler-driven packing optimizations have progressed from fixed-layout heuristics to automatic, layout-flexible strategies that substantially improve utilization for SIMD encoding.
        \item Scale management and bootstrapping placement are tightly coupled challenges. Recent compilers adopt level-aware cost models and region-based scheduling to jointly minimize scale consumption and latency.
        \item Despite these advances, existing approaches still suffer from scalability limitations, such as long compilation times and poor support for complex model architectures like Transformers.
    \end{itemize}
\end{tcolorbox}
\end{center}

\subsection{GPU Optimization}
\label{sec:gpu}

One of the primary obstacles to the widespread adoption of HE is the significant performance gap between homomorphic and plaintext computations. For example, even the most advanced implementations of CNN inference on the CIFAR-10 dataset still require hundreds to thousands of seconds per inference on CPU~\cite{lee2022low,kim2023optimized,benamira2023tt,rovida2024encrypted}. In hybrid schemes that combine HE with MPC, a common strategy to reduce communication overhead is to offload as much linear computation as possible to the HE domain. However, this often increases the computational cost of leveled HE operations, making GPU acceleration essential. Given that GPUs are the dominant compute resource in modern cloud infrastructures, accelerating PPML on GPUs is a natural and promising direction. This section discusses GPU optimization from two perspectives: operation-level acceleration and PPML system-level optimization. We conclude by summarizing the execution time of representative PPML workloads reported in existing GPU-accelerated implementations.

\subsubsection{Operation-Level Optimization}
Early efforts in GPU acceleration for HE primarily focused on polynomial arithmetic, with particular attention to the Number Theoretic Transform (NTT), one of the most computationally intensive operators in HE. Recent works continue to target NTT for optimization.
HE-Booster~\cite{wang2023he} achieves fine-grained NTT by leveraging inter-thread local synchronization to increase overlap and reduce barrier stalls.
\citet{fan2023towards} explores how to efficiently utilize CUDA cores by implementing NTT using FP64 arithmetic units.
TensorFHE~\cite{fan2023tensorfhe} utilizes Tensor Core Units (TCUs) with INT8 components to accelerate NTT computation.
WarpDrive~\cite{fan2025warpdrive} reduces warp-level stalls by concurrently scheduling TCUs and CUDA cores.
Neo~\cite{jiao2025neo} further exploits TCUs by transforming scalar and element-wise multiplications into matrix multiplications using the FP64 components in TCUs.

\textbf{Key switching.}
Key switching (\textsf{KeySwitch}) is a maintenance operation in HE schemes that transforms intermediate ciphertexts, whose associated decryption keys have changed due to prior homomorphic operations, back into a form decryptable by the user's secret key. In practice, \textsf{KeySwitch} often becomes the performance bottleneck in HE workloads.

Prior work has accelerated \textsf{KeySwitch} by optimizing the underlying polynomial arithmetic. For example, Over100x~\cite{jung2021over} adopts an improved hybrid key switching (HKS) algorithm proposed in~\cite{han2020better}.
Phantom~\cite{shen2022carm,yang2024phantom} and WarpDrive enhance HKS further by decomposing the input ciphertexts on demand.
Neo explores GPU acceleration for the state-of-the-art KLSS-based \textsf{KeySwitch} algorithm~\cite{kim2023accelerating}.

\textbf{Bootstrapping.}
As discussed in Sections~\ref{sec:graph} and~\ref{sssec:scale-mgt-boot-plm}, bootstrapping is another essential maintenance operation in HE schemes, particularly for deep workloads such as DNNs. As the most time-consuming component, bootstrapping is commonly used as a benchmark to evaluate the effectiveness in recent studies.

Over100x~\cite{jung2021over} presents the first GPU-accelerated CKKS bootstrapping implementation. Based on the observation that element-wise operations dominate the bootstrapping cost, they propose memory-centric optimizations such as kernel fusion and reordering. TensorFHE introduces packed bootstrapping to improve throughput.
Cheddar~\cite{kim2024cheddar} implements an advanced CKKS bootstrapping algorithm~\cite{bossuat2022bootstrapping} on GPU. Other GPU-accelerated CKKS efforts have not directly optimized bootstrapping itself, instead focusing on accelerating lower-level primitives as discussed above.

Early GPU acceleration efforts for FHEW/TFHE focused on gate bootstrapping.
Zama's TFHE-rs advances this by supporting more practical LUT-based PBS (see Section~\ref{sssec:softmax-app})
Given the relatively small parameters in TFHE (with polynomial lengths not exceeding 1024), TFHE-rs adopts a single-thread-block-per-SM strategy to execute the entire PBS.
\citet{xiao2025gpu} further design a multi-thread-block PBS kernel optimized for small batch sizes.
VeloFHE~\cite{shen2025velofhe} introduces a hybrid four-step NTT algorithm and applies multi-level memory-centric optimizations.

\subsubsection{System-Level Optimization}
At the system level, several studies have explored hardware extensions, scheduling strategies, and heterogeneous optimizations to enhance HE performance on GPUs. GME~\cite{shivdikar2023gme} proposes a custom hardware design based on the AMD GPU architecture to accelerate polynomial operations. BoostCom~\cite{yudha2024boostcom} improves GPU utilization by using CPU multithreading to dispatch GPU kernels for BGV scheme comparison operations. CAT~\cite{li2025cat} proposes a task scheduling strategy that combines kernel segmentation with the CUDA multi-stream technique. Anaheim~\cite{kim2025anaheim} combines GPU with processing-in-memory (PIM) technology to form a heterogeneous accelerator, offloading memory-bound computations to the PIM side for significant performance gains in real workloads.

\subsubsection{PPML Workload}
\label{sssec:ppml-workload}

Commonly used workloads for evaluating GPU-accelerated HE include:
\begin{itemize}
    \item \textbf{HELR}: Homomorphic logistic regression. \textit{HELR-train} refers to the training workload, which executes 32 iterations on 1024 batches of \(14\times14\) grayscale images~\cite{han2019logistic}. Performance is typically reported in the execution time per iteration. 
    \textit{HELR-infer} refers to the inference workload, though its exact definition varies across studies.
    
    \item \textbf{RNN}~\cite{podschwadt2020classification}: Recurrent neural network inference. A single ciphertext encodes 32 embeddings, each of length 128. The benchmark runs 200 iterations of an RNN layer with 128 units. 
    
    \item \textbf{HECNN}: HE-friendly convolution neural network inference. \textit{CryptoNets}~\cite{gilad2016cryptonets} refers to the 5-layer CNN which replaces non-linear activations with squaring functions. Similar lightweight CNNs have also been retrained for TFHE-based inference~\cite{bourse2018fast,ho2024efficient}.
    
    \item \textbf{ResNet}: Convolution neural network inference using ResNet architecture. \textit{ResNet-20/32/56} refers to the inference of a single CIFAR-10 image~\cite{lee2022low}. \textit{ResNet-18} refers to the inference of a single ImageNet image.
\end{itemize}

\begin{table}[ht]
\centering
\begin{adjustbox}{width=\linewidth}
\begin{threeparttable}
\caption{Overview of GPU-accelerated HE frameworks.\tnote{a}} 
\label{tab:gpu_he_overview}
\begin{tabular}{@{}cccccccccc@{}}
\toprule
\multirow{2}{*}{Literature} & \multirow{2}{*}{Year} & \multicolumn{4}{c}{Scheme}              & \multirow{2}{*}{\begin{tabular}[c]{@{}c@{}}Support\\ boot.\end{tabular}} & \multirow{2}{*}{\begin{tabular}[c]{@{}c@{}}PPML\\ benchmark\tnote{b}\end{tabular}} & \multicolumn{2}{c}{Accessibility} \\ \cmidrule(lr){3-6} \cmidrule(l){9-10} 
                            &                       & BFV     & BGV     & CKKS    & TFHE/FHEW &                                                                          &                                                                           & Code            & Exe.           \\ \midrule
Over100x~\cite{jung2021over}& 2021                  & \Circle & \Circle & \CIRCLE & \Circle   & \CIRCLE                                                                  & HELR-train                                                                & \LEFTcircle     & \LEFTcircle    \\
HE-Booster~\cite{wang2023he}& 2023                  & \Circle & \CIRCLE & \CIRCLE & \Circle   & \Circle                                                                  & CryptoNets                                     & \Circle         & \Circle        \\
TensorFHE~\cite{fan2023tensorfhe}& 2023                  & \Circle & \Circle & \CIRCLE & \Circle   & \CIRCLE                                                                  & HELR-train, RNN, ResNet-20                                                & \Circle         & \LEFTcircle    \\
GME~\cite{shivdikar2023gme} & 2023                  & \Circle & \Circle & \CIRCLE & \Circle   & \CIRCLE                                                                  & HELR-train, ResNet-20                                                     & \Circle         & \Circle        \\
BoostCom~\cite{yudha2024boostcom}& 2023                  & \Circle & \CIRCLE & \Circle & \Circle   & \Circle                                                                  & MLP                                                                       & \Circle         & \Circle        \\
Phantom~\cite{shen2022carm,yang2024phantom}& 2024                  & \CIRCLE & \CIRCLE & \CIRCLE & \Circle   & \Circle                                                                  & -                                                                         & \CIRCLE         & \CIRCLE        \\
Cheddar~\cite{kim2024cheddar}& 2024                  & \Circle & \Circle & \CIRCLE & \Circle   & \CIRCLE                                                                  & HELR-train, RNN, ResNet-20                                                & \LEFTcircle     & \CIRCLE        \\
\citet{de2024privacy,deencryptedllm}& 2024                  & \Circle & \Circle & \CIRCLE & \Circle   & \CIRCLE                                                                  & GPT-2                                                                     & \LEFTcircle         & \CIRCLE        \\
ArctyrEX                    & 2025                  & \Circle & \Circle & \Circle & \CIRCLE   & \CIRCLE                                                                  & HELR-infer, HECNN~\cite{bourse2018fast}, AES-128-Dec\tnote{$\dagger$}             & \Circle         & \Circle        \\
\citet{xiao2025gpu}          & 2025                  & \Circle & \Circle & \Circle & \CIRCLE   & \CIRCLE                                                                  & HECNN~\cite{ho2024efficient}                                                & \Circle         & \Circle        \\
CAT~\cite{li2025cat}        & 2025                  & \CIRCLE & \CIRCLE & \CIRCLE & \Circle   & \Circle                                                                  & PDQ\tnote{$\dagger$}                                                            & \CIRCLE         & \CIRCLE        \\
HEonGPU~\cite{ozcan2023homomorphic}& 2025                  & \CIRCLE & \Circle & \CIRCLE & \CIRCLE   & \CIRCLE                                                                  & PIR\tnote{$\dagger$}                                                            & \CIRCLE         & \CIRCLE        \\
WarpDrive~\cite{fan2025warpdrive}& 2025                  & \Circle & \Circle & \CIRCLE & \Circle   & \CIRCLE                                                                  & HELR-train, ResNet-20, AES-128-Dec\tnote{$\dagger$}                        & \Circle         & \Circle        \\
Anaheim~\cite{kim2025anaheim}& 2025                  & \Circle & \Circle & \CIRCLE & \Circle   & \CIRCLE                                                                  & HELR-train, RNN, ResNet-20, ResNet-18                                     & \Circle         & \Circle        \\
Neo~\cite{jiao2025neo}      & 2025                  & \Circle & \Circle & \CIRCLE & \Circle   & \CIRCLE                                                                  & HELR-train, ResNet-20/32/56                                               & \Circle         & \Circle        \\ \bottomrule
\end{tabular}
\begin{tablenotes}
\footnotesize
\item[a] ``\CIRCLE'' = full, ``\LEFTcircle'' = partial/limited, ``\Circle'' = none. For ``Scheme'' columns, this refers to support; for ``Accessibility'' columns, it indicates availability (open-source or binary).
\item[b] ``PPML benchmark'' refers to those defined in Section~\ref{sssec:ppml-workload} or explicitly named in the original papers.
\item[$\dagger$] Not a PPML task: PIR = Private Information Retrieval, PDQ = Private Dataset Query, AES-128-Dec = AES-128 Decryption/Transciphering.
\end{tablenotes}
\end{threeparttable}
\end{adjustbox}
\end{table}

\begin{figure}[!tb]
    \centering
    \includegraphics[width=\linewidth]{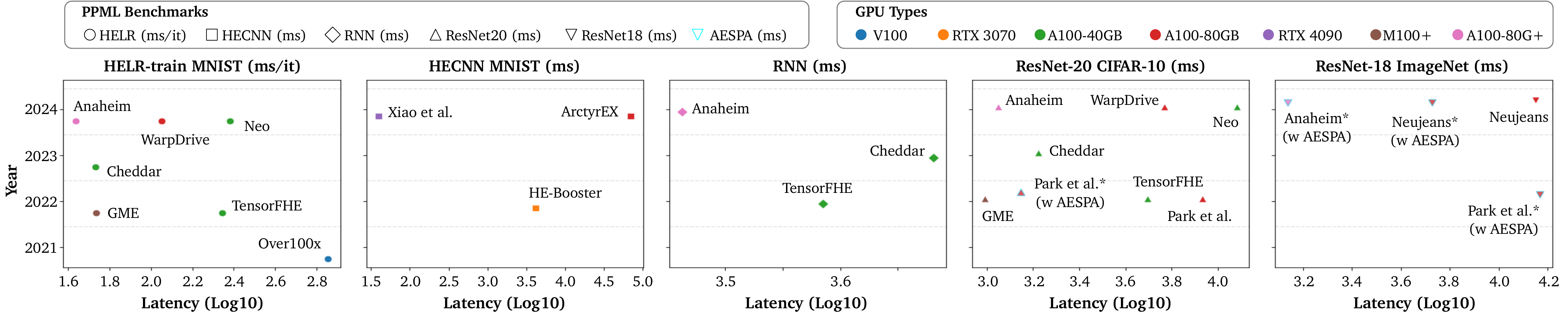}
    \caption{Execution time of GPU-accelerated PPML workloads (as defined in Section~\ref{sssec:ppml-workload}). Note that Neujeans~\cite{ju2024neujeans} and \citet{park2023toward} report only the latency after GPU acceleration. Methods based on AESPA are highlighted. M100+ and A100-80GB+ indicate systems with custom extensions to the original GPUs.
    }
    \label{fig:gpu-report}
\end{figure}

A summary of recent GPU-accelerated HE frameworks is provided in Table~\ref{tab:gpu_he_overview}, including the accessibility of their source code. Performance results for PPML benchmarks are summarized in Figure~\ref{fig:gpu-report}. Currently, only NEXUS~\cite{zhang2024secure} and \citet{de2024privacy} report GPU-accelerated LLM inference using pure HE. In particular, \citet{de2024privacy} evaluates one decoding step of GPT-2 at token position 128, which takes approximately 105 seconds on an A100-80GB GPU. This result is later optimized to 69 seconds~\cite{deencryptedllm}.

\subsection{Implementation and Library}
\label{sec:lib}

We summarize the open-source PPML libraries in Table \ref{tab:libs}. These libraries use a range of cryptographic techniques, including SS, HE, and OT, to ensure that computations are performed securely and without revealing sensitive data. 

\begin{table}[!tb]
    \centering
    \caption{Open-source PPML implementations and libraries.}
    \label{tab:libs}
    \resizebox{\linewidth}{!}{
    \begin{tabular}{ccccccc}
    \toprule
    Library      & Time  & Provided Protocols  &  Front-end Language & Back-end Language &  Repository Link  \\
    \midrule
    HElib \cite{helib}   & 2013 &  HE (BGV, CKKS) &  C++  &  C++ & \url{https://github.com/homenc/HElib} \\
    python-paillier \cite{PythonPaillier} & 2014  &  HE (Paillier)  &  Python & Python & \url{https://github.com/data61/python-paillier} \\
    Pyfhel \cite{ibarrondo2021pyfhel} & 2017 &  HE (BGV, BFV, CKKS)  &  Python & C++ &  \url{https://github.com/ibarrond/Pyfhel} \\
    SEAL \cite{sealcrypto}  & 2018   &  HE (BGV, BFV, CKKS)  &  C++ & C++ & \url{https://github.com/microsoft/SEAL} \\
    SEAL-Python \cite{sealcrypto} & 2019 &  HE (BGV, BFV, CKKS) & Python &  C++ &  \url{https://github.com/Huelse/SEAL-Python} \\
    OpenFHE \cite{openfhe} & 2021 & HE (BGV, BFV, CKKS) & C++, Python & C++ &  \url{https://github.com/openfheorg/openfhe-development} \\
    Lattigo \cite{lattigo} & 2024 & HE (BGV, BFV, CKKS) & Go & Go & \url{https://github.com/tuneinsight/lattigo} \\
    TFHE-rs \cite{zama2024TFHE} & 2024 & HE (TFHE) & C, Rust & Rust & \url{https://github.com/zama-ai/tfhe-rs} \\
    \midrule
    ABY \cite{demmler2015aby} &  2015 &  MPC &  C++  &  C++ & \url{https://github.com/encryptogroup/ABY} \\
    EMP-toolkit \cite{emp-toolkit}  &  2016   & MPC  &  C++  &  C++ & \url{https://github.com/emp-toolkit} \\
    ABY3 \cite{mohassel2018aby3}  &  2018 &  MPC  & C++ &  C++ & \url{https://github.com/ladnir/aby3} \\
    CrypTen \cite{knott2021crypten}  & 2019  &  MPC &  Python  &  Python & \url{https://github.com/facebookresearch/CrypTen} \\
    MOTION \cite{braun2022motion}  & 2022  & MPC  & C++  & C++ & \url{https://github.com/encryptogroup/MOTION}
    \\
    \midrule
    MP-SPDZ \cite{mp-spdz}  & 2016 & MPC, HE & Python  & C++ & \url{https://github.com/data61/MP-SPDZ} \\
    EzPC \cite{chandran2019ezpc}  &  2019 & MPC, HE &   C++  &  C++ & \url{https://github.com/mpc-msri/EzPC} \\
    Secretflow \cite{ma2023secretflow}  & 2022 & MPC, HE & Python & C++ & \url{https://github.com/secretflow/secretflow} \\
    \bottomrule
    \end{tabular}
    }
\end{table}



\section{Discussions}
\label{sec:discussion}

This literature review highlights that while substantial progress has been made, current PPML frameworks still encounter efficiency bottlenecks, underscoring the need for further efforts in this field.
Here, we put forward several discussion points regarding the future development of PPML, which we hope may serve as a starting point for broader research.

\subsection{What are the Challenges and Opportunities of Cross-level Optimization?}

For the future optimization direction, we argue that full-stack, i.e., cross-level optimization holds great promise.
There arises a question: ``Can different levels be decoupled from each other? What are the challenges and unique opportunities when considering cross-level optimizations?''





\textit{\textbf{Discussion point 1: Protocol-model co-optimization.}}
Model-level optimization like quantization usually requires the support of specifically designed protocols.
Protocol-model co-optimization is essential to achieving maximal benefits.
We provide the following cases:
\textbf{1) Quantization cannot directly help PPML.}
Although quantization drastically reduces the computation and memory overhead, it can introduce extra expensive online costs, including bit width extension, truncation, re-quantization, etc.
Existing studies \cite{xu2024privquant,zeng2024eqo,xu2024hequant} propose PPML-friendly quantization algorithms and protocol optimizations like protocol fusion to minimize the overall costs.
\textbf{2) Non-linear layer optimization cannot help reduce total communication.}
Numerous studies focus on ReLU/GeLU pruning for lower online costs; however, they all ignore the importance of reducing total costs for MLaaS, including online and pre-processing costs.
CoPriv \cite{zeng2023copriv} integrates Winograd convolution protocol, ReLU pruning, and layer fusion to cut down the costs of both linear and non-linear layers.
We believe that in the era of MLaaS, overall efficiency matters most for practical deployment.
\textbf{3) Effective exploitation of sparsity is essential for PPML efficiency gain.} Although the attention mechanism exhibits high sparsity in LLMs, direct pruning either leads to accuracy degradation or introduces additional overhead. Recently, MPCache \cite{zeng2025mpcache} designs a static-dynamic pruning algorithm that carefully considers protocol costs, achieving efficient and accurate pruning.
Other AI techniques such as neural architecture search (NAS) \cite{mishra2020delphi,zeng2023mpcvit,zeng2023copriv,xu2024privcirnet}, knowledge distillation (KD) \cite{zeng2023mpcvit,chen2023rna,li2022mpcformer,dhyani2023privit}, and low-rank decomposition can be applied or optimized for PPML.
Nevertheless, we argue that \textbf{extra training is not practical and scalable for LLMs.
Therefore, post-training quantization, sparsity utilization, training-free decomposition, etc, should be prioritized.}

\textit{\textbf{Discussion point 2: Protocol-system co-optimization.}}
System-level optimization for PPML significantly impacts efficiency, yet it often lacks deep consideration of underlying protocols. This leads to sub-optimal performance, highlighting the need for protocol-system co-optimization, especially for \textbf{HE-based protocols}. We highlight the following:
\textbf{1) Compiler-protocol co-design is key for efficient HE.}
Manual FHE programming is challenging due to complex trade-offs in correctness, noise, packing, and latency, underscoring the need for compilers. While compilers improve packing, their efficacy often relies on expertly optimized protocol-level encoding~\cite{ebel2023orion,krastev2024tensor}. Conversely, current compiler-driven packing mostly uses SIMD encoding, neglecting potentially more efficient coefficient encoding. Bridging this gap by integrating protocol insights into compiler frameworks is a major opportunity.
\textbf{2) ML-orient GPU components cannot directly accelerate PPML.}
Modern GPUs with TCUs accelerate unencrypted ML, but their limited precision support poses challenges for the high-precision modular arithmetic required by HE~\cite{fan2023tensorfhe,fan2025warpdrive}. This demands co-design between cryptographic protocols and hardware capabilities. Future protocols may need to be redesigned to leverage TCU features, possibly through novel encoding or approximate operations. Synergistic efforts with model quantization are crucial to fully exploit GPU's new components for encrypted computations.

\subsection{Is There a Way to Compute Non-Linear Layers Directly Using HE without Polynomial Approximation?}

OT-based non-linear layer brings significant communication overhead.
In contrast, HE-based methods are more communication-efficient.
In existing methods, polynomial approximation is necessary for non-linear layer computation.
As mentioned, low-degree polynomials require extra training efforts, while high-degree polynomials lead to much higher online inference overhead.
Hence, there is a natural question: ``Can we directly compute non-linear layers using HE without polynomial approximation?''
This question is still underexplored.


\subsection{Discussions about PPML in the Era of LLMs: Unique Challenges and Our Views}

Finally, we want to discuss PPML in the era of LLMs, which is much more challenging than prior private inference on CNNs and ViTs.
For example, for an input with only 4 tokens, PUMA \cite{dong2023puma} even requires 122 s latency and 0.9 GB communication to generate 1 token, which is impractical for real-world applications.
The key challenges of private LLM inference can be roughly summarized as follows:
\textbf{1) Large-scale linear layers.}Unlike the matrix-vector multiplications commonly used in CNNs, LLMs involve lots of high-dimensional matrix multiplications in the embedding table, Attention layers, and FFN layers. For instance, in the GPT-2 base model, the up projection in the FFN layer requires a matrix multiplication of dimensions $(\#tokens, 768, 3072)$. \textbf{2) Complicated non-linear layers.}
Different from simple ReLU in CNNs, LLMs consist of more complex non-linear functions like Softmax, GeLU, and SiLU, which require expensive exponential, tanh, and division protocols. Due to the large scale of the Transformer, GPT-2 inference requires about $3.9 \times10^6$ point-wise GeLUs.
\textbf{3) Complicate PPML-aware optimizations.}
LLM optimization is much more complex than CNN and ViT, and training-free methods should be prioritized.
As mentioned in MPCache \cite{zeng2025mpcache}, naive KV cache compression fails to deliver the desired efficiency improvements, and thus, PPML-friendly algorithms should be carefully designed.
Furthermore, we also think parameter-efficient fine-tuning (PEFT), like LoRA for PPML-friendly LLM architecture design, can be a promising research direction.
Moreover, due to the massive scale of LLMs, more severe challenges also arise at the system and protocol levels.
\section{Conclusion}

This survey provides a systematic literature review of PPML with structured and meticulous taxonomy, i.e., protocol-level, model-level, and system-level optimization.
This survey also provides qualitative and quantitative comparisons of existing works with technical insights.
Moreover, this survey discusses future research directions and highlights the necessity of integrating optimizations across protocol, model, and system levels, serving as a road map and white book for researchers in the PPML community.


\bibliographystyle{ACM-Reference-Format}
\bibliography{main}


\end{document}